\def\Nat{{\em Nature}}

\def\spose#1{\hbox to 0pt{#1\hss}}
\def\approxlt{\mathrel{\spose{\lower 3pt\hbox{$\sim$}}
         \raise 2.0pt\hbox{$<$}}}
\def\approxgt{\mathrel{\spose{\lower 3pt\hbox{$\sim$}}
         \raise 2.0pt\hbox{$>$}}}

\def\multleft#1{\hbox to size{\vbox {\halign {\lft{##}\cr #1}}\hfill}\par}
\def\multright#1{\hbox to size{\vbox {\halign {\rt{##}\cr #1}}\hfill}\par}

\def\boxit#1{\vbox{\hrule\hbox{\vrule\kern3pt\vbox{\kern3pt
           #1 \kern3pt}\kern3pt\vrule}\hrule}}

\def\cm{{\rm\thinspace cm}}

\def\ergs{{\rm\thinspace ergs}}

\def\s{{\rm\thinspace s}}

\def\chisq{\hbox{$\chi^2$}}

\def\pcmsq{\hbox{$\cm^{-2},$}}

\def\ergpcmsqps{\hbox{$\ergs\cm^{-2}\s^{-1}\,$}}

\def\ergps{\hbox{$\ergs\s^{-1}\,$}}

\def\pcmsq{\hbox{$\cm^{-2}\,$}}

\documentclass[preprint]{aastex}

\received{}
\accepted{}

\slugcomment{Submitted to the Astronomical Journal}

\shorttitle{The {\it Chandra} Large Area Sensitive X-ray Survey }
\shortauthors{Yang, Mushotzky, Steffen et al.}

\begin{document}

\title{The {\it Chandra} Large Area Synoptic X-ray Survey (CLASXS) of the 
Lockman Hole-Northwest: The X-ray Catalog }

\author{Y. Yang\altaffilmark{1,2}, R. F. Mushotzky\altaffilmark{2}, 
A. T. Steffen\altaffilmark{3}, A. J. Barger\altaffilmark{3,4,5}, 
L. L. Cowie\altaffilmark{5}}

\altaffiltext{1}{Department of Astronomy, University of Maryland, 
MD College Park 20742}
\altaffiltext{2}{Laboratory for High Energy Astrophysics, NASA 
Goddard Space Flight Center, Code 662, Greenbelt, MD 20770}
\altaffiltext{3}{Department of Astronomy, University of Wisconsin, 
Madison, WI 53706}
\altaffiltext{4}{Department of Physics and Astronomy, University of 
Hawaii, Honolulu, HI 96822}
\altaffiltext{5}{Institute for Astronomy, University of Hawaii, 
Honolulu, HI 96822}

\begin{abstract}
We present the X-ray catalog and basic results from our {\it Chandra} 
Large Area Synoptic X-ray Survey (CLASXS) of the Lockman Hole-Northwest
field. Our 9 ACIS-I fields cover a contiguous solid angle of 
$\sim 0.4$ deg$^{2}$ and reach fluxes of $5\times 10^{-16}$~\ergpcmsqps 
($0.4-2$~keV) and $3\times 10^{-15}$~\ergpcmsqps ($2-8$~keV). 
Our survey bridges the gap between ultradeep 
pencil-beam surveys, such as the {\it Chandra} Deep Fields (CDFs), 
and shallower, large area surveys, allowing a better probe of 
the X-ray sources that contribute most of the 2--10 keV cosmic X-ray 
background (CXB). We find a total of 525 X-ray point sources and 4 
extended sources. At $\sim 10^{-14}$~\ergpcmsqps ($2-8$~keV), our number 
counts are significantly higher than those of several non-contiguous, 
large area surveys. Such a large difference is an indication of clustering in 
the X-ray sources. On the other hand, the integrated flux from the 
CLASXS field, combined with {\it ASCA} and {\it Chandra} ultradeep 
surveys, is consistent with results from other large area surveys, 
within the variance of the CXB. 

We see spectral evolution in the hardening of the sources at fluxes 
below $10^{-14}$~\ergpcmsqps, which agrees with previous observations 
from {\it Chandra} and {\it XMM-Newton}. About 1/3 of the sources in 
the CLASXS field have multiple observations, allowing variability tests. 
Above $4\times 10^{-14}$~\ergpcmsqps ($0.4-8$~keV), $\sim 60\%$ of the sources 
are variable. We also investigated the spectral variability 
of the variable sources. While most show spectral softening
with increasing flux, or no significant spectral change, there are 
a few sources that show a different trend. 

Four extended sources in CLASXS is consistent with the previously 
measured LogN-LogS of galaxy clusters. Using X-ray spectra and optical colors, 
we argue that 3 of the 4 extended sources are galaxy clusters or galaxy
groups. We report the discovery of a gravitational lensing arc associated 
with one of these sources. Using red sequence and brightest 
cluster galaxy methods, we find that the redshifts of the extended sources 
are in the range $z\sim 0.5-1$. The inferred masses within the Einstein 
radii are consistent with the mass profiles of local groups scaled to the 
same virial radii. 
\end{abstract}


\section{Introduction}

With several ultradeep {\it Chandra} and {\it XMM-Newton} surveys  
reaching flux limits as deep as $f_{2-10~{\rm keV}}\sim 10^{-16}$\ergpcmsqps
(Alexander et al. 2003; Giacconi et al. 2002; Hasinger 2003),
it is now clear that point sources account for almost
all of the X-ray radiation above 2~keV in the universe (Mushotzky et al. 2000;
Moretti et al. 2003). These point sources are believed to be 
primarily active galactic nuclei (AGN), based
on their X-ray luminosities. It has also been found that only
$\sim 30\%$ of hard X-ray selected AGN show broad lines in the optical,
while $>50\%$ appear to be normal galaxies in the optical at the 
sensitivity limits of the current optical spectroscopic follow-up 
(e.g., Barger et al. 2001, 2003). In contrast, the soft X-ray selected 
AGN from {\it ROSAT} surveys are mostly identified as broad-line AGN 
in the optical band (Schmidt et al. 1998).

The ultradeep surveys cover very small solid angles. In the case of 
the {\it Chandra} Deep Fields (CDFs), the combined sky coverage is 
$\sim 0.2$~deg$^{2}$. About 40\% variance between fields is seen in 
the integrated fluxes in the $2-8$~keV band (Cowie et al. 2002), 
likely as a result of the underlying large scale structure. 
To determine the fractional
contribution of point sources to the cosmic X-ray background (CXB)
with enough accuracy, and to understand how the CXB sources trace the
large scale structure, a sufficiently large solid angle is needed.
While very large area surveys exist above $10^{-13}$~\ergpcmsqps (2--8 keV) 
from {\it ASCA} (Akiyama et al. 2003), the data around $10^{-14}$~\ergpcmsqps,
where the point source contribution to the CXB peaks, is limited.

Several intermediate, wide-field, serendipitous {\it Chandra/XMM-Newton} 
surveys (Baldi et al.\ 2002; Kim et al.\ 2004; Harrison et al.\ 2003) 
were designed to increase the solid angle to several degrees at a
$2-8$~keV flux limit of $10^{-14}$~\ergpcmsqps. One of the advantages
of such surveys is that they sample randomly across the sky, so the 
probability of all of them hitting overdense or underdense regions is small. 
This is useful in determining the normalization of LogN-LogS. On the other 
hand, serendipitous surveys suffer from the non-uniform 
observing conditions for each pointing, and in most cases, the pointings 
contain bright sources. The biases introduced by these non-uniformities 
are hard to quantify.  The serendipitous surveys also have little
power in addressing the question of large scale structure traced by X-ray 
selected AGN, due to the small solid angle of each pointing, sparse and 
random positions on the sky, and the non-uniformity of the observations.
With serendipitous surveys, it is also difficult to perform extensive 
optical spectroscopic follow-up observations, which are critical in obtaining 
the redshifts and spectral classifications of the X-ray sources. This 
is due in part to the advent of large format detectors for imaging 
and spectroscopy (like those on the Subaru and Keck telescopes),
which are more efficient at targeting large-area, contiguous X-ray 
surveys, rather than many isolated ACIS-I pointings.

A contiguous, large solid angle survey can compensate for these
disadvantages and bridge the gap
between the ultradeep ``pencil beam'' surveys and the large area
serendipitous surveys in determining both the normalization of the
LogN-LogS and the large scale structure. 

In 2001, we began the {\it Chandra} Large Area Synoptic
X-ray Survey (CLASXS) of the multiwavelength data-rich {\it ISO} 
Lockman Hole-Northwest (LHNW) region. The survey currently covers a 
solid angle of $\sim 0.4$~deg$^{2}$ and is sensitive to a factor of 
$2-3$ below the ``knee'' of the $2-8$~keV LogN-LogS.
Such a choice of solid angle and depth maximizes the detection efficiency
with {\it Chandra}. The large solid angle is important for obtaining  
statistically significant source counts at the ``knee'' of the LogN-LogS 
and to test for variance of the number counts on larger solid angles. 
The choice of solid angle is based on the {\it ASCA} results that the rms
variance of the $2-10$~keV CXB on a scale of 0.5~deg$^{2}$ is $\sim 6\%$
(Kushino et al.\ 2002). The expected variance at the angular scale of 
our field should be less than the uncertainty of the CXB flux.
The uniform nature of the survey allows an unbiased measurement of AGN
clustering. We have shown that the field-to-field variance seen between
the deep fields can be reproduced within the 9 fields in our survey 
(Yang et al.\ 2003; hereafter Yang03). 

Our survey region is covered by the deepest 90 and 170$\mu$m ISOPHOT 
observations (Kawara et al. 2004), as well as abundant multiwavelength 
observations, including the planned {\em Spitzer Space Telescope (SST)} 
Wide-Area Infrared Extragalactic Survey (SWIRE).
We performed extensive optical follow-up observations using
Subaru, CFHT, WIYN, and Keck to obtain multicolor images and spectra
of the X-ray sources (Steffen et al. 2004, hereafter Steffen04).
These observations provide critical information on the redshifts, 
spectroscopic classifications and luminosities of the X-ray sources,  
as well as on the morphologies of
the host galaxies. We emphasize that the subarcsecond spatial 
resolution of {\it Chandra} is critical to allow the unambiguous 
identification of optical counterparts.

We describe our observations and data analysis methods in \S 2 and
present our X-ray catalog in \S 3.  In \S 4, we present the
X-ray spectral properties and variability of the point sources.
In \S 5, we present our analysis of the extended sources and report
on the discovery of a gravitational lensing arc associated with one 
of the clusters.
We summarize the paper in \S 6. A companion paper by Steffen04 presents 
our multiwavelength observations and analysis.  We will present
the spatial correlation functions of the X-ray sources in a subsequent paper.
Throughout this paper, we assume $H_{0}=71$ and a flat universe 
with $\Omega_{M}=0.27$ and $\Omega_{\Lambda} = 0.73$.

\section{Observations and Data Reduction}

\subsection{X-ray Observations}

We surveyed the LHNW field centered at
$\alpha=10^{h}34^{m}$, $\delta=57\arcdeg40\arcmin$
(J2000). The region has the lowest Galactic absorption ($N_{H} \equiv
5.72\times 10^{19}$\pcmsq; Dickey \& Lockman 1990).
All 9 ACIS-I observations were obtained with the standard
configuration. The pointings are separated from each
other by 10\arcmin~(Figure 1). The fields are labeled LHNW1-9 for reference
hereafter. The overlapping of the fields allows a uniform sky coverage,
because the sensitivity of the telescope drops significantly at large 
off-axis angles. Fields LHNW1-3 were observed during April 30th to May 17th
2001, and the rest of the fields were observed during April 29th to May 4th
2002. All fields except LHNW1 have exposure times of
$\sim 40$~ks. LHNW1 is located at the center of the field and has an exposure
time of 73~ks. The observations are summarized in Table 1.

We reduced the data with CIAO v2.3 and the calibration files in CALDB v2.20.
For our spectral analysis, we updated the data reduction with CIAO v3.01 and
CALDB 2.23 to allow the use of CTI corrected calibration files. We followed 
the CIAO analysis threads (available online at http://asc.harvard.edu/ciao/) 
in reducing the data, including the correction of known
aspect problems, CTI problems, and removing high background intervals.
Background flares were found in LHNW3 and LHNW6 and have been removed.
The resulting event lists were rebinned into
$0.4-2$~keV (soft), $2-8$~keV (hard), and $0.4-8$~keV (full) broadband images.
Spectral weighted exposure maps were made for each band for each observation,
using the observation specified bad pixel files.

\subsection{Source detection}

The detection sensitivity of {\it Chandra} 
drops rapidly beyond 6\arcmin~off-axis. For this reason, we overlapped 
our ACIS-I fields so that the sensitivity of the survey would be uniform 
across the field. Since the added signal-to-noise from merging the observations is 
relatively small, we chose to detect sources in each observation 
individually and merge the catalogs, rather than to detect sources 
directly on the merged image. This method certainly loses some 
sensitivity for very dim sources at some locations. However, since our 
major interest is to obtain a uniform sample for statistical and 
follow-up purposes, such a choice is justified. 
The method also simplifies the source flux extraction because the PSF 
information could easily be used. Multiple detections of
sources in independent observations are very useful for checking and 
improving the X-ray positions of the sources. Multiple detections also 
provide an opportunity for measuring the variability of these sources. 

From the various detection tools available, we chose to use the 
{\it wavdetect} tool included in the CIAO package (Freeman et al.\ 2002). 
Since {\it wavdetect} uses a set of scales to optimize the source detection, 
the tool is excellent in separating nearby sources in crowded fields. In 
general, the method also provides better sensitivity than the classical 
``sliding box'' methods. The drawbacks of {\it wavdetect} are that it runs 
rather slowly on large images like the full resolution ACIS-I images, and 
it requires fine-tuning of the parameters.

We ran {\it wavdetect} on the full resolution
images with wavelet scales of $1,\sqrt 2, 2,2\sqrt2,4,4\sqrt2,8$.
Although using larger scale sizes could help to detect very far 
off-axis sources, it is not very useful for our survey, because of the 
overlapping of fields. It also increases the computation time to use a 
large number of scales. We chose to use a significance threshold of 
$10^{-7}$, which translates to a probability of false detection of 0.4 
per ACIS-I field based on Monte Carlo simulation results
(Freeman et al.\ 2002). 

\subsection{Source positions}
Observations performed before 2002-May-02 suffer from an systematic 
aspect offset as large as 2\arcsec~ from an error in
the pipeline software. This systematic error was carefully calibrated 
by the {\it Chandra} X-ray center and corrections are provided. 
For the affected fields, LHNW1, 2, 3, 4, 5, and 7, we corrected
this error following the standard procedures (see {\it Chandra} 
analysis thread 
online at http://asc.harvard.edu/ciao/threads/arcsec\_correction/).

We further matched the small off-axis X-ray positions reported in 
each field from {\it wavdetect} to the optical images. Corrections
were then found to maximize the matches. Such corrections are very small. 
The astrometric improvement also only marginally improved the matching 
between the X-ray catalogs, thanks to the excellent astrometric accuracy
of the instrument. The corrected
X-ray catalogs from each observation were then merged (\S 3.1). A further
absolute astrometric correction was applied to the merged X-ray catalog to 
match to the radio sources in the field. In Figure 2, we show the offsets 
between the corrected positions
and the optical positions for sources in different off-axis angle ranges.
Large offsets often occur when the X-ray source is at large off-axis angles.
Offsets are generally small, with a dispersion of 0.23\arcsec for all the 
sources detected.

\subsection{Source fluxes}
{\it Wavdetect} is excellent at detecting sources, but it is not always the
best method for flux extraction. Three issues could contribute to an
incorrect estimation of source counts in {\it wavdetect}.
First, the flux measurements in
{\it wavdetect} use a monochromatic PSF size,
  which, by default, corresponds to an enclosed energy of 0.393 at the energy
of choice, or the $1\sigma$ integrated volume of a normalized two-dimensional 
Gaussian. Though this parameter is adjustable, larger enclosed energy values 
could cause confusion of close sources. Since the construction of source cells
is carried out by convolving the source image with wavelet functions,
the ``smearing'' effects of the convolution can in general make the source
cell large enough to include most of the source photons, but the fraction
of the flux recovered varies from source to source.  Second, due to the
statistical fluctuations in the source photon distribution,
some sources show multiple peaks in the convolved image. Unless perfect
PSF information is available, randomness should exist in determining which peak
belongs to a single source. This problem is  particularly
severe when
the source is very off-axis and the PSF shape cannot be approximated
by a Gaussian. The third issue is the background determination, a problem
other methods also share. The background in  {\it wavdetect} is obtained
in the immediate neighborhood
of the source. This is useful because of the known large background 
fluctuations.
However, if the background is drawn too close to the source, the PSF wing
would likely  be taken as background. This could result in an
over-subtraction of the background
and lead to underestimated source counts. This effect is seen in a
correlation of
source counts with background density in the {\it wavdetect} results.
All of these issues would lead to an underestimation of source counts. 
This has been noticed in the analysis of the deep {\it Chandra} fields
(Giacconi et al. 2002; Hornschemeier, private communication).

Because of the spectral differences of the sources and the sensitivity
differences
between energy bands, sources detected in one band are not always detected in
another at high significance. There is no simple way within {\it wavdetect}
to provide upper limits for these sources. To obtain the source fluxes or
upper limits in the non-detection band, an alternative
flux extraction method is needed.

For these reasons, we wrote an aperture photometry tool for
source flux  extraction.
The method  uses a simple circular aperture
which matches the size of the PSF. To do this, we first compared the
broadband PSFs derived from our observations with the  PSF size file
provided with CIAO, as described below.

\subsubsection{Broadband PSFs}

Both the PSF library used by the CIAO tool {\it mkpsf} and the 
circularly averged PSFs used by the detection codes (psfsize20010416.fits)
are generated at monochromatic energies
using simulations of the telescope. Spectral weighted average energy is 
usually used for selecting the PSF file for broadband images. 
Since the spectra of the X-ray sources are mostly unknown, an average 
spectrum has to be assumed. Whether
such selected PSFs agree with the observed broadband PSFs needs to 
be tested. We constructed ``average PSFs'' for different off-axis angles 
using sources which have no neighbors within 40\arcsec~ in our
9 observations (Figure 3). It should be noted that these PSFs are inaccurate 
at large scales because the PSF wings, which span more than 
1\arcmin, could not be well determined in these observations. 
The source images from the same off-axis annuli are stacked, 
and the curves-of-growth are constructed. The background regions are
fitted with quadratic forms using nonlinear least-square fits.
To compare with the library PSFs used by {\it wavdetect}, we linearly
interpolated the library PSFs to the off-axis angles and
the spectral weighted averaged energies. To account for the fact that part 
of the PSF wings had been fitted as background in our data, we did the same 
``background fitting'' on the interpolated PSFs. This allows a comparison of
the observed curve with the interpolated PSF. The broadband
PSFs are generally narrower than the interpolated PSFs, except
for one case in the hard band where the off-axis angle is large.

\subsubsection{Aperture photometry}

We performed our flux extractions in the following way. We used 
circular extraction cells, choosing the radius of cells from the PSF library 
at a nominal enclosed energy of $\sim 95\%$ (the true enclosed energy should 
be $>$95\% based on the discussion above) if the cell size is $> 2.5$\arcsec. 
For source close to the aim point, a fixed 2.5\arcsec~ radius was used.
The background is estimated in an annulus region with
an area 4 times as big as the source cell area, with inner radius 5\arcsec~
larger than the source cell radius. To avoid nearby sources being included
in the background region, the background region is divided into 8 
equal-sized segments (Figure 4).
The mean background counts are estimated, excluding the segment which 
contains the
highest number of events. Then the $3\sigma$ Poisson upper limit is derived
using the approximations provided in Gehrels (1986). The background is then
recalculated with only the background segments that contain counts less than
the upper limit. The net counts are obtained by subtracting the background
from the source counts within the source cell. We compare the
obtained net counts with the net counts obtained with {\it wavdetect} 
(Figure 5). While they mostly agree, the source photons derived from our 
method are, on average, higher than those from {\it wavdetect},
especially for low-count sources. The average increases are 4\%, 
7\%, and 8\% for the soft, full, and hard bands, respectively.  We 
hand-checked the sources with large discrepancies from the two methods, 
and we found our estimates to be more reliable.

\subsubsection{Exposure time and flux conversion}

The prerequisite for using exposure maps is that the effective area is only
weakly dependent on energy. This is not the case for our broadband images,
where the effective area changes rapidly with energy. Using exposure maps blindly,
even the spectrally weighted ones, will inevitably introduce large errors in
the resulting fluxes. However, the vignetting (the positional changes of 
sensitivity) is less sensitive to energy. In other words, if we normalize 
the exposure maps obtained at different energies to the aim points, then 
the differences between such ``normalized exposure maps'' are very small. 

Based on this fact, we use the exposure maps only to correct for vignetting 
and compute the flux conversion at the aim point using spectral modeling. 
We first make full resolution spectrally weighted exposure maps (using 
monochromatic maps do not change the results significantly). 
For each source, the exposure map is convolved with the PSF generated using 
{\it mkpsf} and normalized to the exposure time at the aim point. This is 
the effective exposure time if the source is at the aim point. 

The conversion factor is then obtained at the aim point by assuming the 
source has a galactic, absorbed, single power-law spectrum. The power-law 
index is calculated using the hardness ratio of each source, defined as 
$HR \equiv C_{hard}/C_{soft}$, where $C_{hard}$ and $C_{soft}$ are the 
count rates in the hard and soft bands. XSPEC was used in computing the 
conversion from $HR$ to $\Gamma$ and for calculating the conversions. 
The degradation of quantum efficiency during the flight of the observatory 
has been accounted for using the standard procedure.

\section{Catalog}

\subsection{Merging Catalogs}

We first merged the three band catalogs. We used a $3\sigma$ error 
ellipse from the {\it wavdetect} output as the identification cell.
Flux extraction was then performed on all entries in the merged
catalogs in all bands using the best position of the sources.
We compared the three band catalogs with the optical
catalog to find the astrometric corrections for each observation, 
as described in \S 2.3. The nine catalogs were then merged. The 
fluxes of the sources with more than one detection in the 9 fields 
were taken from the observation in which the effective area of
the source was the largest, except for those sources with more than 2
detections having normalized areas $>80\%$, where we took the averaged
flux. We visually checked the final catalog to ensure the correctness 
of the merging process. The final catalog contains 525 sources.

The distribution of the source off-axis angles in the merged catalog is
shown in Figure 6. It can be seen that most of the sources fall within the
$<6$\arcmin~ range. Figure 7 shows the distribution of sources with multiple
detections. About 1/3 of the sources have more than one observation.

\subsection{ Column-by-column description of the catalog}

We present the final catalog in two tables (Table 2a and Table 2b). 
In Table 2a, we list the source positions, fluxes, and hardness ratios. 
In Table 2b, we list the source net counts, effective exposures, 
and detection information.

{\bf Table 2a: Basic properties}

{\bf Column 1}: Source number used in the catalog. The numbers 
correspond to ascending order of right ascension.

{\bf Column 2}: Source name follows the IAU convention and should 
read CXCCLASXS, plus the name given in the table.

{\bf Columns 3 -- 4}: The X-ray position, corrected 
for the aspect errors of the telescope, if applicable, and for the
general astrometric solution by comparing with the optical images 
(see \S 2.3). For sources with multiple detections in the three bands 
and the 9 observations, the best position is taken.

{\bf  Columns 5 -- 6}: Statistical error of the X-ray position quoted 
from the {\it wavdetect} lists.

{\bf  Columns 7 -- 9 }: X-ray fluxes in the soft, hard, and full bands 
in units of $10^{-15}$~\ergpcmsqps.  If a source is detected in multiple 
observations, and if there are more than one observation in which the source 
effective area is more than 80\% of the effective area at the aim point, 
then the mean flux is used; otherwise, the flux from the observation that 
has the largest effective area is used. The errors quoted are the $1\sigma$ 
upper and lower limits, using the approximations from Gehrels (1986). 
For any source detected in one band but with a very weak signal in another, 
the background subtracted flux could be negative. In this case, only the 
upper limit is quoted.

{\bf  Columns 10}: Hardness ratio as defined in \S 2.4.3. The upper or
lower limit is listed for a source with no net extracted photons in 
the soft or hard bands.

{\bf Table 2b: Additional Properties}

{\bf Column 1}: Source number.

{\bf Columns 2--4}: Net counts in the soft, hard, and full bands. If 
a source is detected in multiple observations, then the observation in 
which the source has the largest effective area (see Column 8) is used. 
As in Table 2a, for the sources with negative counts, only the upper 
limits are listed.

{\bf Columns 5--7}: Effective exposure time in each of the three energy 
bands from the exposure map.

{\bf Columns 8--9}: Detection information. Column 8 is the LHNW 
field number where the source has the largest effective area. 
Column 9 lists the LHNW field numbers (each digit represents a field 
number) in which the source has been detected in at least one of the 
three bands. Sources with multiple detections are necessary for the
detection of variability (see \S 5.2).

\section{Results}
\subsection{Number counts}
\subsubsection{Incompleteness and Eddington Bias}

Incompleteness can be caused by energy or positional dependence of
the sensitivity of X-ray telescopes. Because the spectrum of a source carries
important information on the physical nature of the source itself,
sources of different spectra are usually categorized as different types.
The energy dependent sensitivity acts like a filter in selecting
``hard'' and ``soft'' types of X-ray sources. The
soft band detected sources always contain more soft spectrum objects than
the hard band detected sources and {\it vice versa}. Unless the fraction of
each type remains constant for all fluxes (which we now know is not true),
the energy dependent incompleteness cannot be easily corrected. This issue
is  very important in interpreting the fraction of different types of objects
in flux limited surveys. It is desirable to obtain number counts for
each type of source, but it is hard to do that for the CXB sources,
because the spectra are hard to determine. For our medium deep survey, it
is sensible to follow tradition and only discuss the number counts in
energy bands.

The positional dependent incompleteness is caused by vignetting
and aberration of the X-ray optics. The vignetting
causes the effective area to drop with off-axis angle, and the
aberration makes the off-axis PSF larger so that it includes a larger number
of background events in the source cell. The net effect is that the
sensitivity of source detection drops with increasing
off-axis angle. The sky area is therefore flux dependent.

These effects can be investigated via Monte Carlo simulations.
We first generated
background images using observations of fields \#1 and \#4, which represent
the 70~ks and 40~ks exposures. Point sources are removed from the images, and
the holes left in the images are filled by sampling the local background. 
Random sources are generated uniformly on the background
images. The fluxes of the sources are generated by randomly sampling a
complete subset of the combined {\it Chandra} Deep Fields catalog
(Alexander et al. 2003). The subset contains only sources with hard band
fluxes $> 5 \times 10^{-16}$\ergpcmsqps and effective exposure times 
$> 200$~ks.
The input fluxes are converted to
on-axis counts assuming power-law spectra with $\Gamma=1.4$. The exposure map
for each image is consulted to find the vignetting effect at the source
location, and the normalized exposure is multiplied by the true counts
to obtain the ``observed'' net counts. Only sources with more than 3 counts
are used in the simulation to avoid adding too many undetectable dim sources
to the background. We use the CIAO tool {\it mkpsf} to generate realistic
source shapes at the source positions and energies. The PSF is then sampled
to have the same number of photons as in the source. We chose to use
{\it mkpsf} instead of using the {\it Chandra} simulator MARX because we find
the PSF library used by  {\it mkpsf} better resembles sources
at large off-axis angles. The number density of the sources is chosen
to be 2 times higher than the observed density to increase the number of
simulated sources without affecting detections. We ran 100 simulations
on the 2 fields and the three energy bands and detected the sources using
{\it wavdetect} with identical parameter settings to those we used in 
preparing the
observed catalog. Because of the large computation
time, the number of simulations that can be done is limited.

We then compared the output catalogs with the input source catalogs.
Because of the small size of the simulation, the completeness within 4\arcsec~
is not well determined, and a 5\% uncertainty exists in the 
determined fractions.
Fortunately, the PSF effect is small at such small off-axis angles.
For a given flux threshold, the fraction of source detections
drops monotonically with off-axis angle. This relation is fitted between
4\arcmin~ and 10\arcmin~ with a linear least-squares fit. The 95\% complete
off-axis angle limit is then taken from the interpolation of the fit.
The resulting flux thresholds versus off-axis radii are shown for
the three bands and the two exposure times in Figure 8. We note that 
at large off-axis angles, the sensitivity drops rapidly. This is due to 
the choice of wavelet scales. When the largest
scale used becomes smaller than the PSF size of the source, 
{\it wavdetect} is no longer sensitive. This effect, however, is not 
important for our observations, 
because most of the sources of interest are within 6\arcmin~off-axis, 
thanks to the overlapping of fields. With these curves, we are
able to make threshold maps for the combined catalogs of all three bands.
Sources at very large off-axis angles are excluded from the study
of the LogN-LogS. The combined solid angle versus
flux thresholds is shown in Figure 9.

The Poisson fluctuations in the source fluxes could result in an
overestimation of number counts close to the detection limits.
This is known as the Eddington Bias. The effect
depends both on the slope of the LogN-LogS and the level of fluctuation.
For the CLASXS field, the detection threshold is below the ``knee'' of the
LogN-LogS, and the Eddington bias is relatively small. We corrected this
bias using the method described in Viklinin et al. (1995).
In Figure 10, we compare the average input flux with the average output
flux at different off-axis angles from the simulations.
For the soft band, the correction is only important below
$2 \times 10^{-15}$~\ergpcmsqps. For the hard band, the correction is
important below $8 \times 10^{-15}$~\ergpcmsqps. We fit flux--flux
curves in Figure 10 for the different off-axis angles with fourth
order polynomials and correct the source fluxes in the
observed catalog using these fits.
   
\subsubsection{Number counts}
Sources are selected by consulting the threshold map
at the source
positions and including only those with Eddington bias corrected
fluxes higher than the
threshold map values. Sources very far off-axis are excluded from the analysis.
With these selections, we used a total of 310 and 235 sources in the soft
and hard bands, respectively, to construct the LogN-LogS. The 
cumulative LogN-LogS relations are computed using the formula
\begin{equation}
  N(>S) = \sum_{S_{i}>S}{ \frac{1}{\Omega(S_{i})}} \,,
\end{equation}
where $\Omega$ is the complete solid angle.
We show the results in Figure 11 in the soft
and hard bands with $1\sigma$ Poisson errors.
The differential LogN-LogS for the two bands are shown
in Figure 12, which are calculated using the formula
\begin{equation}
  \frac{dN}{dS} = \sum{ \frac{1}{\Omega_{i} \Delta S}} \,,
\end{equation}
in units of deg$^{-2}$ per $10^{-15}$~\ergpcmsqps.
We fit the resulting differential number counts
with single or broken power laws in the form of
\begin{equation}
\frac{dN}{dS} = n_{0}(\frac{S}{10^{-14}})^{-\alpha}
\end{equation}
using error weighted least-square fits. Since our survey best samples
the ``knee'' of the LogN-LogS, the slope of the power-laws are not
well constrained due to the
lack of data points both far above and below the ``knee''. On the other hand,
$n_{0}$ is better determined, to within $1\%$.

For the soft band, we fit the number counts between
$10^{-15}$ and $10^{-14}$~\ergpcmsqps with a power law. We find the 
best-fit parameters to be $\alpha = 1.7 \pm 0.2$ and 
$n_{0} = 12.49 \pm 0.02$. The slope is in good agreement
with previous observations, such as the {\it Chandra} Deep Field-North 
($1.6 \pm 0.1$, Brandt et al.\ 2001),
SSA13 ($1.7 \pm 0.2$, Mushotzky et al.\ 2000), and the compiled wide fields
from {\it Chandra}, {\it XMM-Newton, ROSAT}, and {\it ASCA} ($1.60^{+0.02}_{-0.03}$,
Moretti et al.\ 2003; hereafter, Moretti03). The normalization also
shows excellent agreement with the compiled results from the 
large area survey of Moretti03, which has an effective solid angle at 
$10^{-14}$~\ergpcmsqps larger than that of CLASXS.
Above $10^{-14}$~\ergpcmsqps, the slope steepens, but the
fluctuations in the number counts make it difficult to find a reasonable fit.
However, the LogN-LogS is apparently consistent with a slope of
$\alpha = 2.5$, shown as the dotted line at these fluxes.

Similarly, we model the hard band number counts with a broken
power law and obtain the following best-fit parameters.
For $S>10^{-14}$~\ergpcmsqps,
$\alpha = 2.4 \pm 0.6$ and $n_{0} = 45.6 \pm 0.5$; for
$3\times 10^{-15} < S < 2 \times 10^{-14}$~\ergpcmsqps,
$\alpha = 1.65 \pm 0.4$ and  $n_{0} = 38.1 \pm 0.2$.
For comparison, we also plot the best-fit cumulative LogN-LogS 
from Moretti03 and differential LogN-LogS
from the SEXSI fields (Harrison et al.\ 2003)
and from Cowie et al.\ (2002). At fluxes below $8 \times 10^{-15}$\ergpcmsqps, 
the differential LogN-LogS for all the fields agrees within the errors.  
The difference in the total counts at a flux limit between the CLASXS 
field and the Moretti03 fields is also small.
An apparent difference is seen around $10^{-14}$~\ergpcmsqps: the total
counts at $10^{-14}$\ergpcmsqps are $\sim 70\%$ higher than those from
Moretti03. This is significant at greater than the $3\sigma$ level.

The integrated flux between
$3 \times 10^{-15}$ and $8 \times 10^{-14}$\ergpcmsqps
from the LogN-LogS is
$(1.2 \pm 0.1) \times 10^{-11}$~ergs~cm$^{-2}$~s$^{-1}$~deg$^{-2}$.
This is $\sim 20\%$ higher than that from Moretti03 and SEXSI in the
same flux range.  Since there is little difference in the
number counts between the CLASXS fields and the other large solid 
angle surveys at
fluxes lower than $8 \times 10^{-15}$\ergpcmsqps, we should expect little
difference below the survey limit on the same angular scales. If 
integrated to lower
fluxes, and including the integration from {\it ASCA} above
$8\times 10^{-14}$\ergpcmsqps, the fractional difference between the
CLASXS field and the other large solid angle surveys can be reduced
to $\sim 10\%$ without considering the possible biases. This difference is
higher than expected from the variance in the CXB from {\it ASCA} observations
but is consistent with recent observations with
{\it RXTE/PCA}, where a 7\% variance is seen among several
$\sim 1$  deg$^{2}$ fields (Revnitsev et al. 2003).

The uncertainty of the hard CXB itself is $\sim 10-15\%$.
The differences in the integrated point source
fluxes from various large fields are within this uncertainty. In
terms of the true contribution from point sources to the CXB, a field with a
solid angle of $\sim 0.3$~deg$^{2}$ seems to be  large enough to be
representative.

The detection of the large variance in the hard band source counts 
around $10^{-14}$~\ergpcmsqps on our survey scale seems to indicate
that the sources that emerge at this flux are more clustered
on the sky than the soft band selected sources.
These sources could account for most of the cosmic variance observed.
This issue will be discussed in a separate paper.

\subsection{Spectral properties and variability of the CXB sources}

\subsubsection{Hardness ratio}
We employ a hardness ratio (\S 2.4.3) to quantify statistically 
the spectra of the CXB sources in our field. Figure 13 shows the 
distribution of hardness ratio versus full band flux.
We have also marked the hypothetical photon indices ($\Gamma$), 
assuming the hardness
ratio change is purely due to the slope change of a single power-law 
spectrum. At fluxes $> 3 \times 10^{-14}$~\ergpcmsqps, most sources 
cluster around $\Gamma \sim 1.7$. At lower fluxes, the
hardness ratio distribution scatter increases and the relative number 
of hard sources increases. Below $3\times 10^{-15}$~\ergpcmsqps, the
data show a paucity of hard sources. This is a selection effect caused 
by the sensitivity in the hard band being lower than in the soft band 
for most spectra. We stacked the sources in flux bins and calculated 
the hardness ratios of the stacked spectra.
Figure 14 shows the stacked hardness ratios from both the CLASXS
$>10^{-14}$~\ergpcmsqps sample and the combined CDFs 
$>10^{-15}$~\ergpcmsqps sample. The flux thresholds are chosen
to avoid selection effects caused by the sensitivity differences 
between the soft and hard bands.  It is apparent that the results 
from our data and those from the CDFs agree well.

The spectral flattening at low fluxes has been observed by several authors
(e.g., Mushotzky et al. 2000; Tozzi et al. 2001;
Piconcelli et al. 2003; Alexander et al. 2003) with observations of 
different depths. Spectral analyses with {\it XMM-Newton} observations 
indicate that such
a flattening is mainly caused by absorption. These obscured AGN must
dominate the population around the ``knee'' of the LogN-LogS to account for
the flat spectrum of the CXB. Since most of the {\it XMM-Newton} 
spectral observations
have reached a few times $10^{-14}$~\ergpcmsqps (Piconcelli et al. 2002),
and the mean spectrum at this threshold is still too soft compared 
with that of the CXB,
a sharp increase of obscuration or a change of spectral shape 
at a flux $\sim 10^{-14}$~\ergpcmsqps is
inevitable. Such a sharp change is seen in the change of hardness ratio
in our wide-field sample.

\subsubsection{Variability}

X-ray variability is an important factor in distinguishing AGN from
starburst galaxies.
Almost all AGN vary in X-rays, except those sources which are Compton
thick. Alexander et al. (2001) showed that only a small fraction
of the optically faint X-ray sources vary. Possible explanations
could be that a large fraction of the optically faint sources
are Compton thick, or that the amplitude of variation of the optically
faint sources is much lower than that of the broad and/or narrow-line 
AGN at the same flux thresholds.

We examine the variability of sources that have been detected 
in more than one of our observations. Since the observations 
were taken in two groups, separated by about one year, and each 
group of observations were taken within a few days
(see Table 1), we are able to test variability on timescales of 
days and/or one year, depending on the location of the source.

For timing analyses with low counts per bin, the usual $\chi^{2}$ 
statistic is inadequate. We use the $C$-statistic (Cash 1979) in 
testing the significance of variability. Cash (1979) showed that 
the $C$-statistic (a reduced form of likelihood ratio) written as
\begin{equation}
\Delta C = -2 \sum_{i=1}^{N}{ [n_{i} ln(e_{i}) - e_{i} -n_{i} ln(n_{i})+n_{i}]}
\end{equation}
is asymptotic to a $\chi^{2}$ distribution
with $N-1$ degrees of freedom, where $n_{i}$ is the observed counts in
the $ith$ sample, $e_{i}$ is the expected counts in that sample, 
and $N$ is the total number of samples used.
We restricted the sample for the variability test to sources with
expected counts greater than 10 in all observations.
The null hypothesis rejection probability was chosen to be 0.01.

A total of 168 sources were tested for variability, of which 42 sources are
significantly variable and 28 sources show variability on timescales of days.
There are 29, 16, and 30 variable sources detected in the soft, hard, 
and full bands, respectively. Figure 15 shows the light-curves of the 
sources that were tested to be variable in any of the three energy bands. 
In the top panel of Figure 16, we show the fraction of
variable sources detected versus flux. 
Between $4-8\times 10^{-14}$~\ergpcmsqps,
70\% of the sources tested show variability. This fraction drops dramatically
as the flux decreases and, at $10^{-14}$~\ergpcmsqps, reaches below 20\%. 
This is at least in part due to the selection effect that larger 
variability is needed at lower fluxes to
make the test significant. At fluxes above $8\times 10^{-14}$~\ergpcmsqps, 
only one of the four sources tested (25\%) was found to be variable.

Following Nandra et al.\ (1997), we define the magnitude of variability 
as the ``excess variance'', the error subtracted $rms$ variance
\begin{equation}
\sigma_{rms}^{2} = \frac{1}{N \mu^2} \sum_{i=1}^{N} 
[(f_{i}-\mu)^{2}-\sigma_{i}^{2}] \,,
\end{equation}
where $f_{i}$ is the flux in each observation, $\mu$ is the mean of 
the fluxes, and $\sigma_{i}$ is the Poisson error of the flux. 
By assuming the same power density spectrum of X-ray variability for 
all AGN, $\sigma_{rms}^{2}$ can be used as a good indicator of whether 
the variability exceeds the Poisson noise. It has been found that there 
exists a good anti-correlation between $\sigma_{rms}^{2}$ and AGN 
luminosity (Nandra et al.\ 1997) in local AGN samples.

In Figure 17, we show the excess variance of sources that had been 
detected to be variable versus X-ray flux in the three energy bands.
At high fluxes, the average $\sigma_{rms}^{2}$ is significantly lower 
than at lower fluxes. As mentioned above, variability is harder to 
detect for low flux sources, unless the source is more variable 
than that of the brighter sources, so this bias could explain why 
there are very few low flux, low variability sources in the plot.
In addition, the sources we detect to be variable are generally soft.
This is consistent with the observation from the CDFs
that optically faint sources (most of which are hard spectrum AGN) 
are less variable (Alexander et al. 2001).  

\subsubsection{Spectral variability}

Very little is known about the spectral variability of the sources 
that contribute the most to the CXB due to a lack of data. Spectral 
variability is seen in about half of the well-studied brighter sources, 
with a general trend of softening of the $2-10$~keV spectra with 
increasing source intensity. But a counterexample is NGC7469, where 
the spectrum flattens when the source flux
increases (Barr 1986). The variability could be accounted for either with
a change in the relative normalization of the different spectral components
or by variation in the absorption.

In Figure 18, we show hardness ratios versus full band fluxes
for the variable sources. While most of the
sources show either no clear spectral variability, or a trend of spectral
softening with increasing flux, there are a number of sources that clearly
become harder with increasing flux. There are also a few sources that 
exhibit a mixed trend. On average, these sources tend to have softer 
spectra with increasing flux.

\section{Extended Sources}

\subsection{Detection}

We searched the $0.4-2$~keV images of each observation for extended sources
using the {\it vtpdetect} tool provided in CIAO. The method uses Voronoi
tessellation and percolation to identify dense regions above Poisson noise.
This method performs best on smooth overdense regions but could confuse crowded
point sources.  We chose to use a threshold scale factor of 0.8 and a maximum
probability of false detection of $10^{-6}$ and to restrict the number of 
events per source to $>30$. We used default values for the rest of the 
parameters. This choice of parameters maximizes the detection
of low surface brightness sources at high significance. We visually examined 
the source list to screen out apparent blended point sources.
The candidates were then selected by comparing the 99\% PSF
radius with the equivalent radius of the source region, and only sources
with a PSF ratio (defined as $\sqrt(A/\pi) / r_{99}$, where $A$ is the area 
of the source region reported by {\it vtpdetect}, and $r_{99}$ is the 
99\% PSF radius at the off-axis angle) higher than 10 were considered 
extended (Table 3).

Four sources were found to be significantly extended, and
all but Source 3 have an off-axis angle of $<5$\arcmin~
in the X-ray observations. Source 3 is at an off-axis angle of 8.4\arcmin.
With the X-ray image alone, one could not rule out the source being a 
blend of point sources. However, a bright gravitational lensing arc 
found in the optical image (see \S 5.6) at the X-ray peak
makes it very likely that the X-ray emission is associated with a 
cluster. Considering the
non-uniformity of the detection due to vignetting and PSF effects, 
the number counts for extended sources above
$3.7\times 10^{-15}$\ergpcmsqps are roughly $>10$~deg$^{-2}$. This 
agrees with the LogN-LogS of clusters at these fluxes found in the 
CDFs (Bauer et al. 2002). It is interesting to note that all 4 extended 
sources are found on only two of the overlapping ACIS-I fields in the 
north of the LHNW region.

\subsection{Optical imaging observations}

Optical observations are describe in detail in Steffen04. Optical images 
were obtained with Suprime-Cam on Subaru (in $B$, $V$, $R$, $I$, $z$\arcmin~) 
and with the CFH12K camera on the Canada-France-Hawaii Telescope 
(CFHT; in $R$, $B$, and CFHT $z$\arcmin). The $2\sigma$ limits in 
$B$, $V$, $R$, $I$, and $z$\arcmin~ are 27.8, 27.5, 27.9, 26.4, and 26.2. 

X-ray contours overlaid on $R$ band optical images are shown in 
Figure 19. We examined the number counts of galaxies within circular 
cells with fixed radii of 0.5\arcmin. At a threshold of $R < 24$, a 
total of 19, 28, 18, and 9 galaxies were found within the cells 
centered at the X-ray peaks of each extended source. 
Because a star is found at 0.6\arcmin~ south-east
of the X-ray peak of Source 2, the galaxy counts could be underestimated. 
Compared with the expected 6.7 galaxies per cell obtained from the whole 
field, the overdensities of galaxies in Sources~1 and 2 are $>3\sigma$, 
while the overdensity of galaxies in Source~3 is $\sim 3\sigma$. 
Source~4 does not show significant clustering of galaxies in the $R$ 
band image. Sources~1 and 2 are very close to each other, 
with a separation of $\sim 2$\arcmin. The closeness and the elongated 
morphology of the two sources suggest that they are undergoing 
interactions. Source~3 is extended along the east-west direction 
with multiple peaks. All 4 sources show bright elliptical galaxies 
at the X-ray peaks.

\subsection{X-ray spectra}

We extract very coarse spectra (grouped to $>15$ counts per bin to 
allow the use of the $\chisq$ statistic) and attempt to constrain 
the properties of the clusters. We fit the data with a simple MEKAL 
model in XSPEC (v11.2), with a fixed abundance of 0.3 of the solar 
value and a fixed Galactic absorption. We restrict the spectral fitting 
to within $0.5-5$~keV, because the signal-to-noise ratio is poor 
outside of this range.

The source extraction and background regions of Sources~1 and 2 are shown
in Figure 20. The regions avoid the point sources between the two clusters.
The spectra are shown in Figure 21. For Source~1, we found the best fit
to be kT$=1.4^{+1.0}_{-0.2}$~keV and $a=0.5^{+0.2}_{-0.2}$, with a reduced
$\chisq = 8.8$ for 9 degrees of freedom. This agrees with the redshift
estimates using the optical data (Table 4). Fitting the same model to the
spectrum of Source~2 with the redshift fixed to $z=0.5$ yields
kT$=3.1^{+13.5}_{-1.6}$~keV, with a reduced $\chisq=4.6$ for 9 degrees of 
freedom.  The constraint on the temperature is poor, but the probability 
that the temperature of Source~2 is significantly different than that of 
Source~1 is low. This can be seen in Figure 22, where the joint probability 
contour of the temperature from the two sources is shown. The confidence 
level for the two sources having different temperatures is only $2\sigma$. 
Combining the two data sets and fixing $z=0.5$, we find 
kT$=1.7^{+2.2}_{-0.5}$~keV.

The spectrum of Source~3 shown in Figure 23 was extracted from a circular
region with radius 36\arcsec. The background was extracted from an annulus
with inner radius 36\arcsec~ and outer radius 60\arcsec.
The data cannot constrain the model very well, but
a simple fit with an absorbed power-law shows that the spectrum is
very soft with photon index $\Gamma = 2.6$ and reduced $\chisq=7.2$
for 7 degrees of freedom. Fitting with a MEKAL model and assuming a
redshift of $z=1.1$ (see \S 5.5), we obtain a  temperature of 
$2.3^{+1.1}_{-0.8}$ with reduced $\chisq = 0.77$. The temperature is 
insensitive to the redshift between $z=0.4-1.4$. The fact that the MEKAL 
model fits the data better makes it less likely that Source~3 is a 
blend of several point sources.

With only 30.7 net counts, it is impossible to model the spectrum for 
Source~4. However, the source has very few counts above 2~keV, 
indicating that the temperature should be low if the source is at 
$z > 0.4$, as implied from the optical data.

The virial masses of the extended sources can be roughly estimated 
using the best-fit $M-L$ relation (Finoguenov et al.\ 2001), 
$M_{500} = 2.45 \times 10^{13}~{\rm T}^{1.87}$, where $M_{500}$
is the mass within a radius where the overdensity is 500. The results
are shown in Table 5. All of the sources belong to low mass clusters 
or groups, and this result is not very sensitive to the redshift because 
of the very soft spectra.

\subsection{Angular sizes}

The angular sizes of the sources were quantified by the widths of the 
radial profiles. We fitted the radial profiles of the sources with 
integrated 2-D Gaussian curves, which describe the low S/N ratio data 
reasonably well. We constructed the cumulative counts as a function of 
off-source radius (curve-of-growth). Exposure maps were applied to 
correct for vignetting. Nearby point sources were removed and replaced 
with background noise. The background regions were selected visually and 
fitted with a quadratic form plus a constant. The curves-of-growth were
then normalized to the best-fit backgrounds. The normalized 
curve-of-growth for each source is shown in Figure 24. This left only 
one parameter to be determined---the widths of the curves. The best-fit 
$1\sigma$ radii are listed in Table 5.

\subsection{Redshifts}

We infer the redshifts of the extended sources using the red sequence 
method, as well as the brightest cluster galaxy (BCG) method. 
Based on observations of clusters, there is usually a population 
of early-type galaxies which follow a color-magnitude relation (red 
sequence). This relation changes with redshift in a predictable way, 
such that a robust two-color photometric redshift can be obtained 
(Gladders \& Yee 2000). Color-magnitude plots of the sources within 
0.5\arcmin~ of the X-ray centers are shown for each extended source 
in Figure 25. Red sequences can be clearly seen in Sources 1, 2, and 4.  
By comparing with the models from Yee \& Gladders (2001),
we can estimate the redshifts for these three extended sources (Table 4). 
Source 3 does not show a clear red sequence.

BCGs are often used as distance indicators, because they have
almost constant luminosity (Humasom, Mayall, \& Sandage 1956). 
One of the difficulties in applying this method is that with 
optical images alone, it is hard to distinguish between the 
background and the cluster members, unless a density peak can 
be clearly determined. In our case, this is less worrisome because
bright spheroidal/lenticular galaxies are found at the X-ray peaks 
of all of the extended sources.
This clearly associates these galaxies with the clusters.
Furthermore, these galaxies are also the brightest early-type galaxies 
in the regions where X-ray emission is significant. Following 
Postman \& Lauer (1995), we fit the radial profile
of each of the BCGs to obtain the magnitudes within angular radius $r_{m}$ 
and the slope of the profile ($\alpha \equiv d log L_{m} /d log r |_{r_{m}}$, 
where $L_{m}$ is the luminosity within $r_{m}$. We eye examine the 
profile so that nearby galaxies are not included in the aperture.
K correction is performed using $K_{R} = 2.5 log_{10} (1+0.96z)$.  
We then solve the redshifts for each galaxy by assuming the cosmological
parameters in \S1. The resulting redshifts are listed in Table 4. 
 
While it appears that the BCG method produces higher redshifts than the 
red sequence method, the differences are not significant, given the 
large uncertainties in both methods. The redshifts of Sources~1 and 2 
also agree with the spectral fitting results from the X-ray data.

The X-ray luminosities of the extended sources are listed in Table 5,
assuming the red sequence-determined redshifts except for Source 3, where 
BCG redshift is adopted. Within errors, the 
temperatures and luminosities of the sources agree with the scaling 
law found in high-redshift X-ray clusters (Ettori et al.\ 2003), but 
the constraint is weak.

\subsection{Discovery of a gravitational lensing arc}

We have found a gravitational lensing arc close to Source 3 (Figure 26).
The arc has an angular radius of $\sim 6$\arcsec~.  A bright 
spheroidal galaxy is clearly associated with the arc.
A possible counter arc is seen connecting to the west of the bright 
galaxy but is not fully resolved. With $B, V, R, I,$ and $z^{\prime}$ 
observations, we can estimate photometric redshifts for the cD galaxy 
and the arc using the publicly available photometric redshift code 
Hyperz (Bolzonella et al.\ 2000). We find photometric redshifts for 
the cD galaxy and the arc of $z=0.45$ and $z=1.7$, respectively. The 
redshift of the galaxy is slightly different than the redshift of the 
cluster obtained using the BCG method. From our experience, one often 
needs at least 7 colors to obtain a secure photometric redshift. The 
redshift estimates therefore need verification. We now discuss the 
estimated lensing properties.

If the source is at $z_{src}\sim 0.45$ and the arc is at 
$z_{arc}\sim 1.7$, then we can estimate the mass within
the Einstein radius (reasonably approximated by the radius of the arc) as
\begin{equation}
M(\theta < \theta_{E}) = 1.1 \times 10^{14} 
(\frac{\theta}{30\arcsec})^{2} (\frac{D_{LS}D_{L}}{D_{S}}) M_{\sun} \,,
\end{equation}
where $D_{L}$, $D_{S}$, and  $D_{LS}$ are, respectively, the angular 
diameter distances (in units of Gpc) of the lens, source, and the 
distance between the lens and source. With $z_{src}\sim 0.45$ and 
$z_{arc}\sim 1.7$, we obtain
$M(\theta < \theta_{E}) \sim 3.3 \times 10^{12}$ M$_{\sun}$. 
We compare this mass with what would be expected if the source were a 
group of galaxies at $z=0.45$, assuming the mass profiles are self-similar.  
By fixing the redshift, the X-ray spectra yield a temperature of 2.2~keV. 
The virial radius is roughly $r_{500} = 0.63 \times \sqrt(kT) = 955$~kpc 
(Finoguenov et al.\ 2001), where $r_{500}$ is defined as the radius 
within which the overdensity is 500.
The size of the arc at $z \sim 0.45$ is $r_{arc} \sim 37$~kpc $=0.036r_{500}$. 
Comparing with the mass profiles of NGC2563, NGC4325, and NGC2300 
(Mushotzky et al.\ 2003), the mass inside the Einstein radius agrees 
very well with that of a group of galaxies. The virial mass of the group 
can then be estimated to be $\sim 1.2 \times 10^{14}$~M$_{\sun}$  
(Finoguenov et al.\ 2001).

If the cluster is at $z \sim 0.7$, as implied from the BCG method, and if 
the best-fit temperature $kT = 0.23$~keV is assumed, then we can search 
for the best redshift of the lensed galaxy, so that the mass within the
Einstein radius agrees with the mass profile of groups. We find
that if the lensed galaxy is at $z = 1.8$, then the mass within
the Einstein radius is $M(\theta < \theta_{E}) \sim 3 \times 
10^{12}M_{\sun}$, which fits the mass profile of groups.
  
If the redshift estimate is correct, then the arc system is very similar 
to the one discovered in the {\it ROSAT} deep survey of the Lockman Hole 
(Hasinger et al.\ 1998b). High-redshift gravitational lensing arcs are 
rare objects so far observed. However, since our large area survey is 
very similar in sky area and depth to the {\it ROSAT} Deep Survey, and 
since both have produced a detection of a strong arc, the probability 
of detection seems high.  Larger area surveys of X-ray selected clusters 
of galaxies with deep optical follow-up would help to determine the 
probability of detection. Such observations should put useful constraints 
on $\Omega_{m}$ and on the density of galaxies at high redshifts (Cooray 1999). 

It is interesting to note that all four of our clusters may have
redshifts $z\sim 0.4 - 0.5$ and are located within a region of only 
$\sim 20$\arcmin~ at the north-east corner of our field. This 
corresponds to a comoving radius of $\sim 5$~Mpc. The implications 
of such large scale structure on the CXB need to be investigated further.

\section{Summary}

In this paper, we presented our CLASXS X-ray catalog. Our survey covers
a $\sim 0.4$ deg$^{2}$ contiguous area in an uniform manner and 
reaches fluxes of $5\times 10^{-16}$~\ergpcmsqps in the $0.4-2$~keV 
band and $3\times 10^{-15}$~\ergpcmsqps in the $2-8$~keV band. We
found a total of 525 point sources and 4 extended sources. We
summarize our results as follows.

\noindent 
1) The number counts in the $0.4-2$~keV band agree very well
with other large area surveys. On the other hand, the number counts 
in the $2-8$~keV band deviate significantly from other large area 
surveys at the ``knee'' of the LogN-LogS, possibly as a result of
the underlying large scale structure.  The total $2-8$~keV band flux
agrees with the observed CXB flux within the observed variance of
the CXB, indicating that the true normalization of the CXB can be 
determined using fields with solid angles $\sim 0.3-0.4$~deg$^{2}$.

\noindent
2) The hardness ratios of the sources in the CLASXS field show a 
significant change at $f_{2-8~{\rm keV}}\sim 10^{-14}$~\ergpcmsqps, 
which bridges the range sampled by previous studies and confirms
the results found in deep {\it Chandra/XMM-Newton} surveys. About 
60\% of the sources with full band fluxes $>4 \times 10^{-14}$ show 
significant variability, while the fraction drops dramatically with 
decreasing flux, at least partly due to selection effects. 
Most sources show no change of hardness ratio or anti-correlation 
with flux. But some sources show a positive correlation or mixed trends. 

\noindent
3) We report on the X-ray and multicolor analysis of four extended 
sources. We argue that the sources are likely low mass clusters or 
groups at redshifts $\sim 0.5$. Two of the clusters are probably 
interacting or merging. 

\noindent
4) We report on the discovery of a gravitational lensing arc. The 
lensing cluster is consistent with being at a redshift of $z=0.45-0.7$.

\acknowledgements
We thank Dr. Jean Swank for the suggestion of C-statistics, 
Ann Horschemeier for discussions on {\it wavdetect}, and an anonymous
referee for comments that helped to improve the paper. Y. Yang also 
thanks Dr. Chris Reynolds for his encouragement and help during this 
thesis work. This paper would not have been possible without the 
excellent work of the supporting staff at CXC and the superb instrument 
of the {\it Chandra X-ray Observatory}.

This project was partially funded under the IDS program of R. Mushotzky.
We also gratefully acknowledge support from NASA's National Space
Grant College and Fellowship Program and the Wisconsin
Space Grant Consortium (A.~T.~S.), CXC grant GO2-3191A (A.~J.~B.), 
NSF grants AST-0084847 and AST-0239425 (A.~J.~B.) 
and AST-0084816 (L.~L.~C.), and the University of Wisconsin Research 
Committee with funds granted by the Wisconsin Alumni Research Foundation,
the Alfred P. Sloan Foundation, and the David and Lucile
Packard Foundation (A.~J.~B.). This paper is part of Y. Yang's PhD 
thesis work at the University of Maryland.

\vfil\eject\clearpage
\begin{deluxetable}{ccccccc}
\tablenum{1}
\tabletypesize{\footnotesize}
\tablewidth{0pt}
\tablecaption{Observation Summary}
\tablecolumns{5} \tablehead{\colhead{Target Name} & \colhead{$\alpha_{2000}$}
& \colhead{$\delta_{2000}$} & \colhead{Obs ID} & \colhead{Sequence \#} &
\colhead{Observation date} & \colhead{Exposure\tablenotemark{a}}} \startdata
LHNW1 & 10 34 00.24 & +57 46 10.6 & 1698 & 900057 & 05/17/01 18:29:38 
& 72.97 ks \\
LHNW2 & 10 33 19.82 & +57 37 13.8 & 1699 & 900058 & 04/30/01 10:59:38 
& 40.74 ks \\
LHNW3 & 10 34 36.12 & +57 37 10.9 & 1697 & 900056 & 05/16/01 12:46:50 
& 43.72 ks \\
LHNW4 & 10 32 04.20 & +57 37 15.6 & 3345 & 900184 & 04/29/02 03:23:45 
& 38.47 ks \\
LHNW5 & 10 34 00.31 & +57 28 15.6 & 3346 & 900185 & 04/30/02 02:03:59 
& 38.21 ks \\
LHNW6 & 10 33 20.28 & +57 55 15.2 & 3343 & 900182 & 05/03/02 09:11:41 
& 34.04 ks \\
LHNW7 & 10 32 44.23 & +57 46 15.2 & 3344 & 900183 & 05/01/02 20:03:06 
& 38.54 ks \\
LHNW8 & 10 34 36.26 & +57 55 15.6 & 3347 & 900186 & 05/02/02 14:16:27 & 38.46 
ks \\
LHNW9 & 10 35 14.28 & +57 46 15.2 & 3348 & 900187 & 05/04/02 11:01:47 
& 39.52 ks \\
\enddata
\tablenotetext{a}{Total good time with dead time correction.}
\end{deluxetable}

\vfil\eject\clearpage

\vfil\eject\clearpage

\vfil\eject\clearpage

\begin{deluxetable}{ccccccccc}
\tablenum{3}
\tabletypesize{\footnotesize}
\tablewidth{0pt}
\tablecaption{Extended Sources}
\tablecolumns{9} \tablehead{
\colhead{Source \#} &
\colhead{$\alpha_{2000}$} &
\colhead{$\delta_{2000}$} &
\colhead{$\Delta \alpha$ (\arcsec)} &
\colhead{$\Delta \delta$ (\arcsec)} &
\colhead{$\theta$ (\arcmin)\tablenotemark{a}} &
\colhead{PSF ratio\tablenotemark{b}} &
\colhead{Net Counts\tablenotemark{c}} &
\colhead{Field\tablenotemark{d}}}
\startdata

    1 &         10 35 25.4 &  +57 50 48 &     2.628 &     1.044 & 
4.786 &   37.89 & $100.1\pm 11$ & 9 \\
    2 &         10 35 13.4 &  +57 50 17 &     3.312 &     1.476 & 
4.029 &   39.33 & $87.7 \pm 11$ & 9 \\
    3 &         10 35 37.9 &  +57 57 15 &     3.060 &     1.476 & 
8.422 &   19.09 & $77.0 \pm 10$ & 8 \\
    4 &         10 34 30.8 &  +57 59 12 &     3.132 &     2.196 & 
4.016 &   16.69 & $30.7 \pm 6.6$ & 8 \\

\enddata
\tablenotetext{a}{Off-axis angle in the field the source is detected. }
\tablenotetext{b}{Defined as $\sqrt(A/\pi) / r_{99}$, where $A$ is 
the area of the
source region from the {\it vtpdetect} report and $r_{99}$ is the 99\% PSF
radius at the off-axis angle.}
\tablenotetext{c}{Net counts reported by {\it vtpdetect}}
\tablenotetext{d}{LHNW field number where the source has the smallest off-axis
angle.}
\end{deluxetable}

\begin{deluxetable}{cccc}
\tablenum{4}
\tabletypesize{\footnotesize}
\tablewidth{0pt}
\tablecaption{Redshift estimates for the extended Sources}
\tablecolumns{4} \tablehead{
\colhead{Source \#} &
\colhead{$z_{RS}$} &
\colhead{$z_{BCG}$ }   &
\colhead{$z_{X-ray}$}}
\startdata

1 & 0.50  &  $0.58^{+0.08}_{-0.08}$ & $0.5 \pm 0.2$ \\
2 & 0.50  &  $0.73^{+0.09}_{-0.08}$ & $0.5 \pm 0.2$ \\
3 & ..... &  $0.73^{+0.09}_{-0.08}$ & .....         \\
4 & 0.45  &  $0.45^{+0.06}_{-0.05}$ & .....         \\

\enddata
\end{deluxetable}


\begin{deluxetable}{ccccccc}
\tablenum{5}
\tabletypesize{\footnotesize}
\tablewidth{0pt}
\tablecaption{Properties of the extended sources}
\tablecolumns{7} \tablehead{
\colhead{Source \#} &
\colhead{$z_{fix}$} &
\colhead{kT\tablenotemark{a}} &
\colhead{$M_{500}$\tablenotemark{b}} &
\colhead{core radius (\arcsec) } &
\colhead{$f_{0.5-8keV}$\tablenotemark{c}}   &
\colhead{$L_{bol}$}\tablenotemark{d}}
\startdata

1 & .50 & $1.4^{+0.8}_{-0.4}$ & $0.45^{+.0.61}_{-.21}$ & 12.9 & 1.6  & 2.2 \\
2 & .50 & $3.1^{+6.5}_{-1.4}$ & $ 2.0^{+15.}_{-1.4}$ & 17.0 & 1.2  & 1.5 \\
3 & .73 & $2.3^{+1.0}_{-0.9}$ & $1.2^{+1.2}_{-.64}$ & 14.7 & 1.5  &  5.1 \\
4 & .45 & 1.0 (fixed)         & .24                 & 11.8 & .42  & .45 \\

\enddata
\tablenotetext{a}{Listed are single parameter $1\sigma$ errors.}
\tablenotetext{b}{unit: $10^{14}$ M$_{\sun}$}
\tablenotetext{c}{Unit: $10^{-14}$\ergpcmsqps.}
\tablenotetext{d}{Unit: $10^{43}$\ergps.}
\end{deluxetable}

\vfil\eject\clearpage

\figcaption[fig1.ps] {Layout of the 9 ACIS-I pointings. Gray scale
map shows the adaptively smoothed full band image. The exposure
maps are added (light gray) to outline the ACIS-I
fields. Fields are separated by 10\arcmin~from
each other. The field numbers (LHNW1-9) are shown at the center of
each ACIS-I field.
\label{fig:layout}}

\figcaption[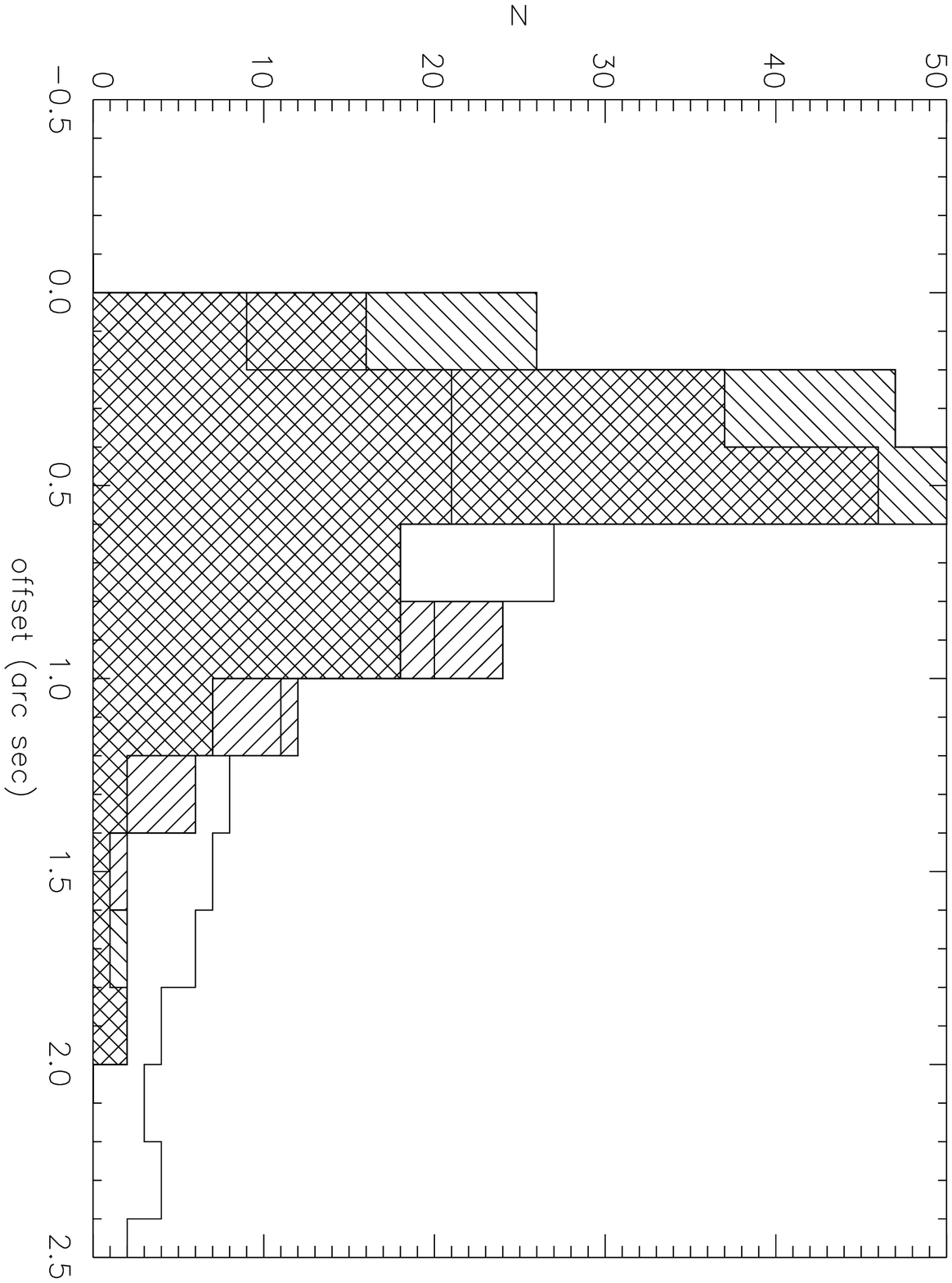] {Offset between the corrected X-ray positions
and the optical positions. The 135\arcdeg~ and 45\arcdeg~, as measured from 
positive x-axis, shaded histogram represents sources with best positions 
within 4\arcmin~ and between 4\arcmin~ 
and 6\arcmin~ of the optical axis, respectively. The unshaded histogram shows 
the sources with off-axis angles $>6\arcmin$.
\label{fig:astrometry}}

\figcaption[fig3.ps] {Broadband PSFs obtained from our observations
{\em (solid lines)\/} compared with the monochromatic PSFs from the PSF 
library {\em (dashed lines)\/}. Both the observed broadband PSFs and 
the monochromatic PSFs are normalized to the wing. From the narrowest 
to the broadest, each broadband PSF is constructed
within each of the off-axis angle intervals 0\arcmin--4\arcmin,
4\arcmin--6\arcmin, 6\arcmin--7\arcmin, 7\arcmin--8\arcmin,
8\arcmin--9\arcmin, 9\arcmin--12\arcmin. The library PSFs are
taken at the midpoints of these off-axis intervals.
(a) Soft band PSF vs. 0.91~keV library PSF; (b) hard band PSF vs. 4.2~keV
library PSF.
\label{fig:im_grey}}

\figcaption[fig4.ps] {Examples of the source and background regions used
  in the flux extraction. The smaller circle is the source region. The
  background regions are shown as segments of an annulus. Segments with
  counts below $3\sigma$ of the mean are used in the final background
  estimation and are marked with `X' symbols. (a) An isolated source;
(b) a source with a close neighbor.
\label{fig:im_grey}}

\figcaption[fig5.ps] {Comparison of net counts from {\it wavdetect} and our
aperture photometry (marked as XPHOTO) for the (a) soft, (b) hard, and 
(c) full bands.
\label{fig:wavdetect_vs_xphoto}}

\figcaption[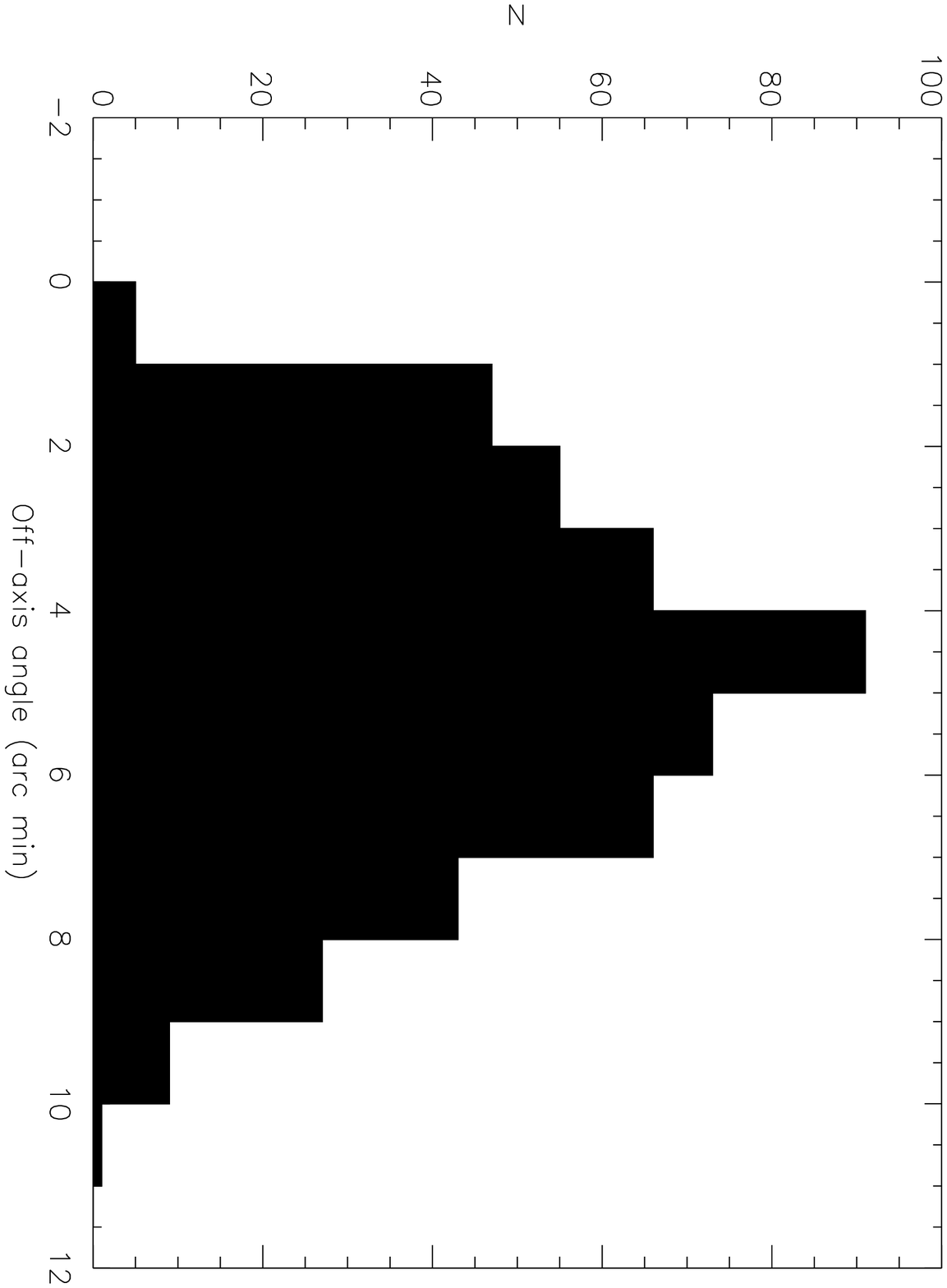] {Distribution of off-axis angles of the best positions.
\label{fig:offaxis_dist}}

\figcaption[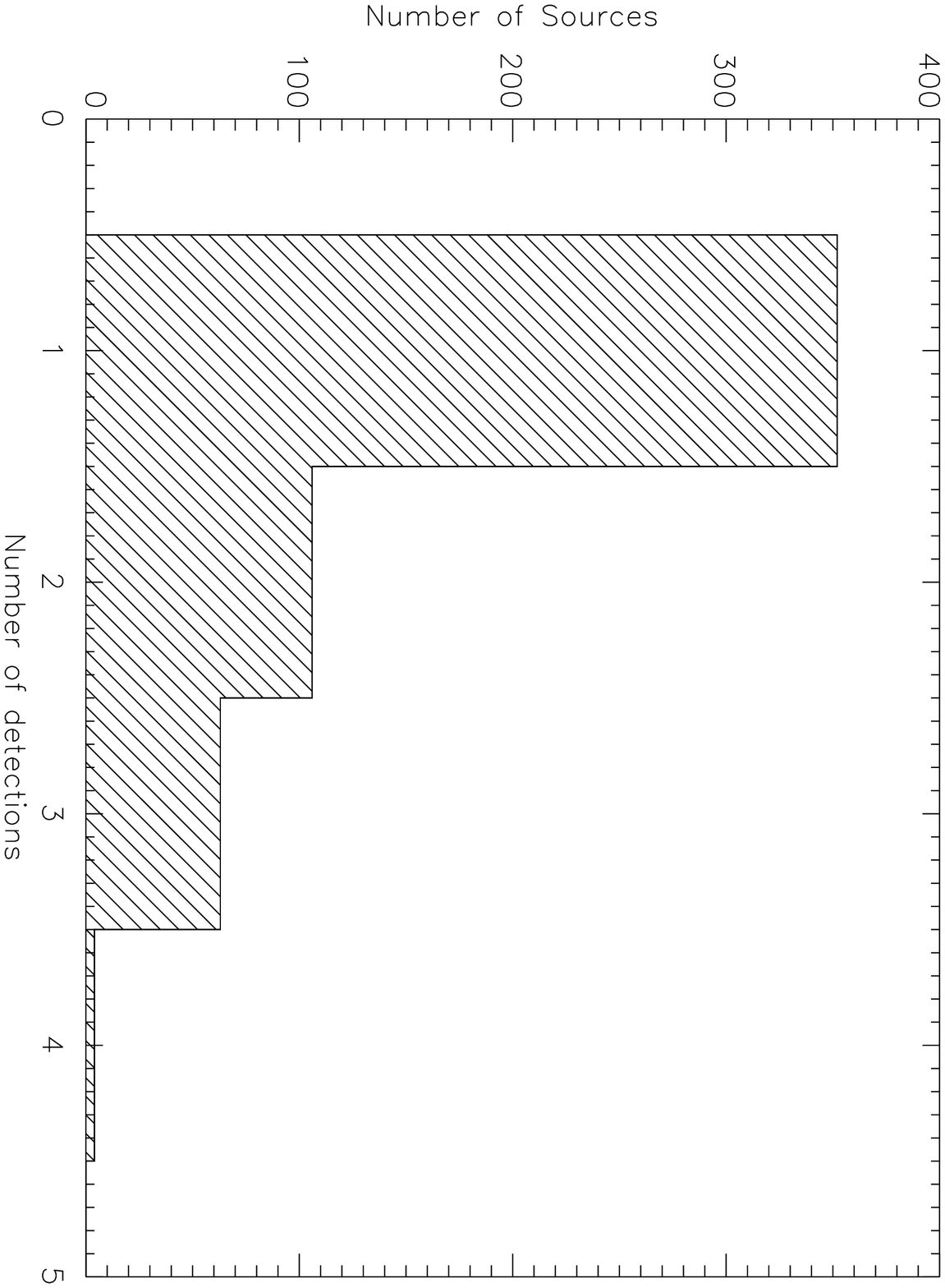] {Distribution of multiple detections.
\label{fig:multiple_detect}}

\figcaption[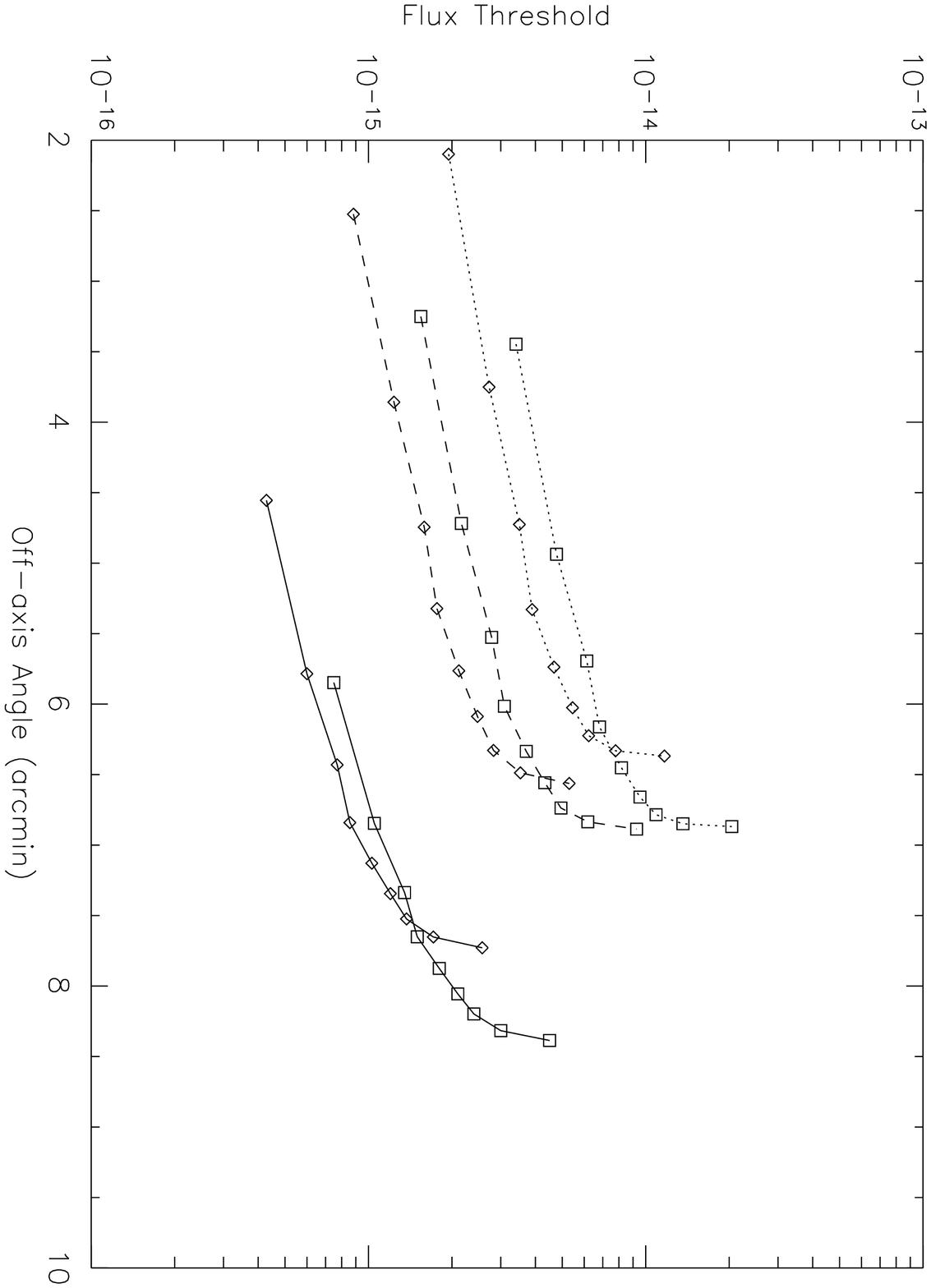] {Thresholds for 95\% completeness vs. off-axis angle.
Sources with flux and off-axis angle combinations above these curves are
complete. Solid, dashed, and dotted lines represent the threshold curves
for the soft, hard, and full bands, respectively. Squares represent the 
70~ks exposure and diamonds the 40~ks exposure.
\label{fig:wavdetect_vs_xphoto}}

\figcaption[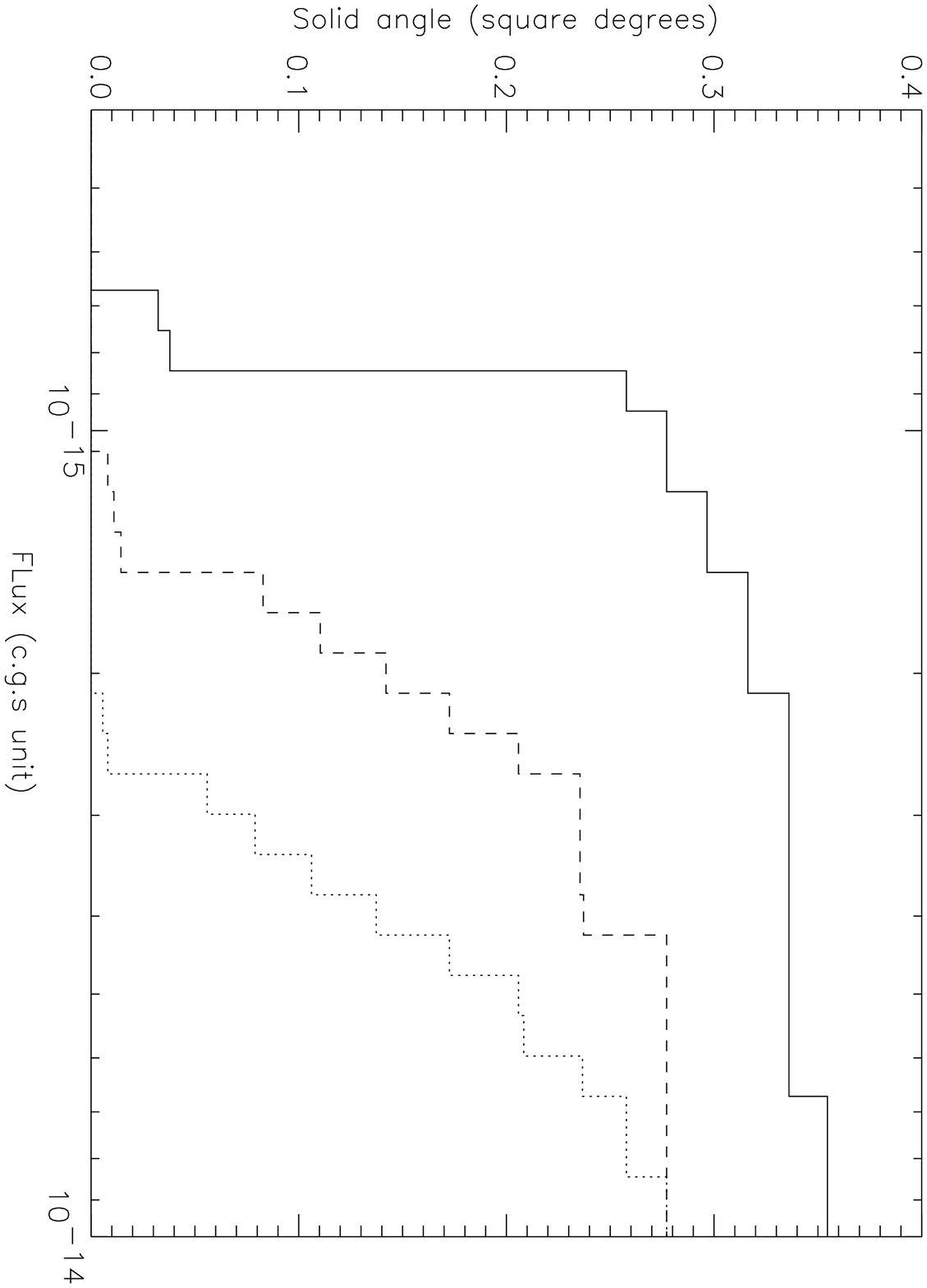] {Survey effective solid angle vs. flux.
Soft band {\em (solid line)}; full band {\em (dashed line)}; 
hard band {\em (dotted line)}.}

\figcaption[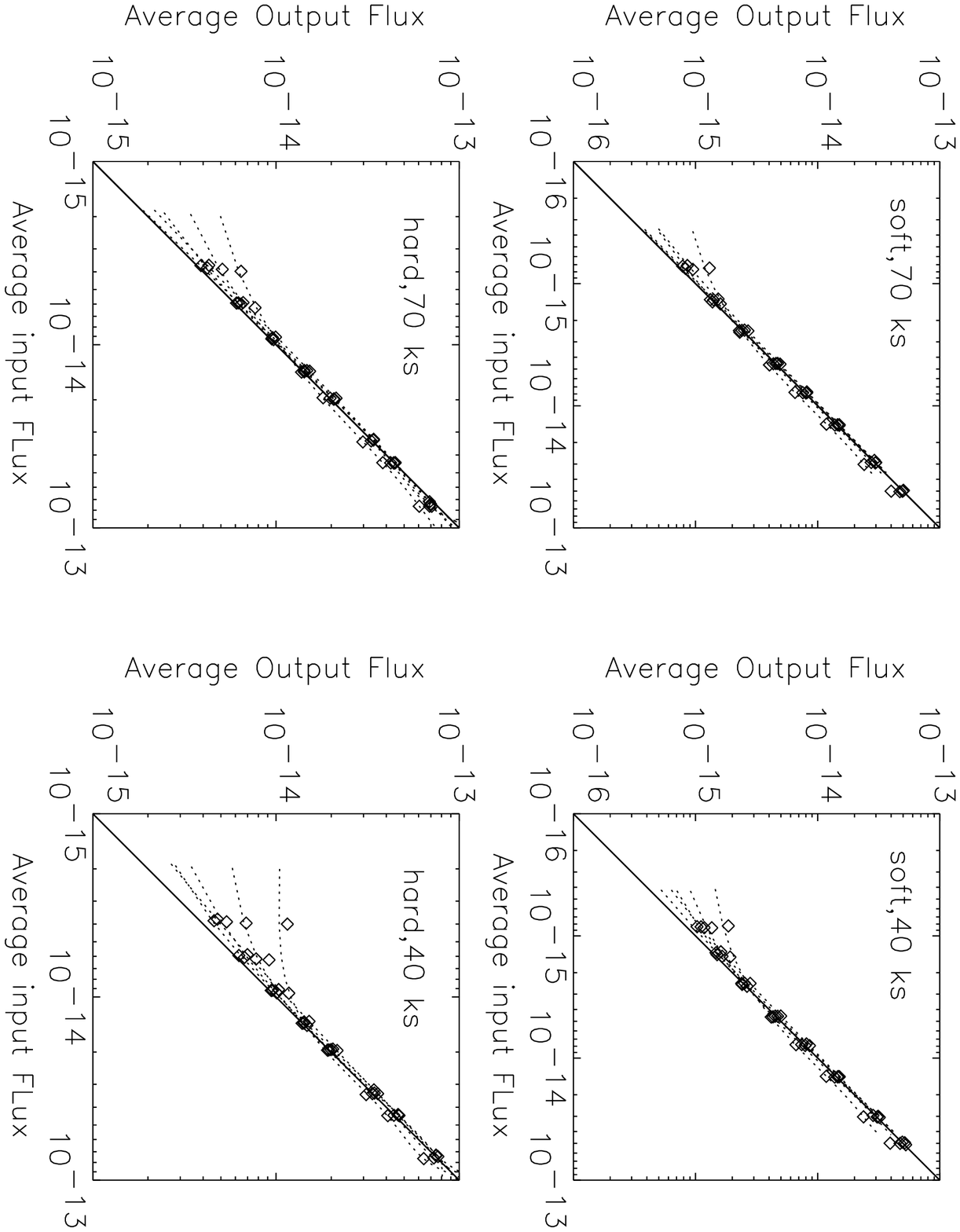] {Average output fluxes from {\it wavdetect}
  vs. average input fluxes for the simulated sources at a set
of off-axis angles {\em (diamonds)}. The Eddington bias is seen in the
overestimates of output flux at low fluxes. The bias also increases
at large off-axis angles. The best fit of the biases are shown as dotted
lines for off-axis angle intervals 0\arcmin--2.5\arcmin,
2.5\arcmin--4\arcmin, 4\arcmin--6\arcmin, 6\arcmin--8\arcmin, and $>8$\arcmin.
\label{fig:Eddington_bias }}

\figcaption[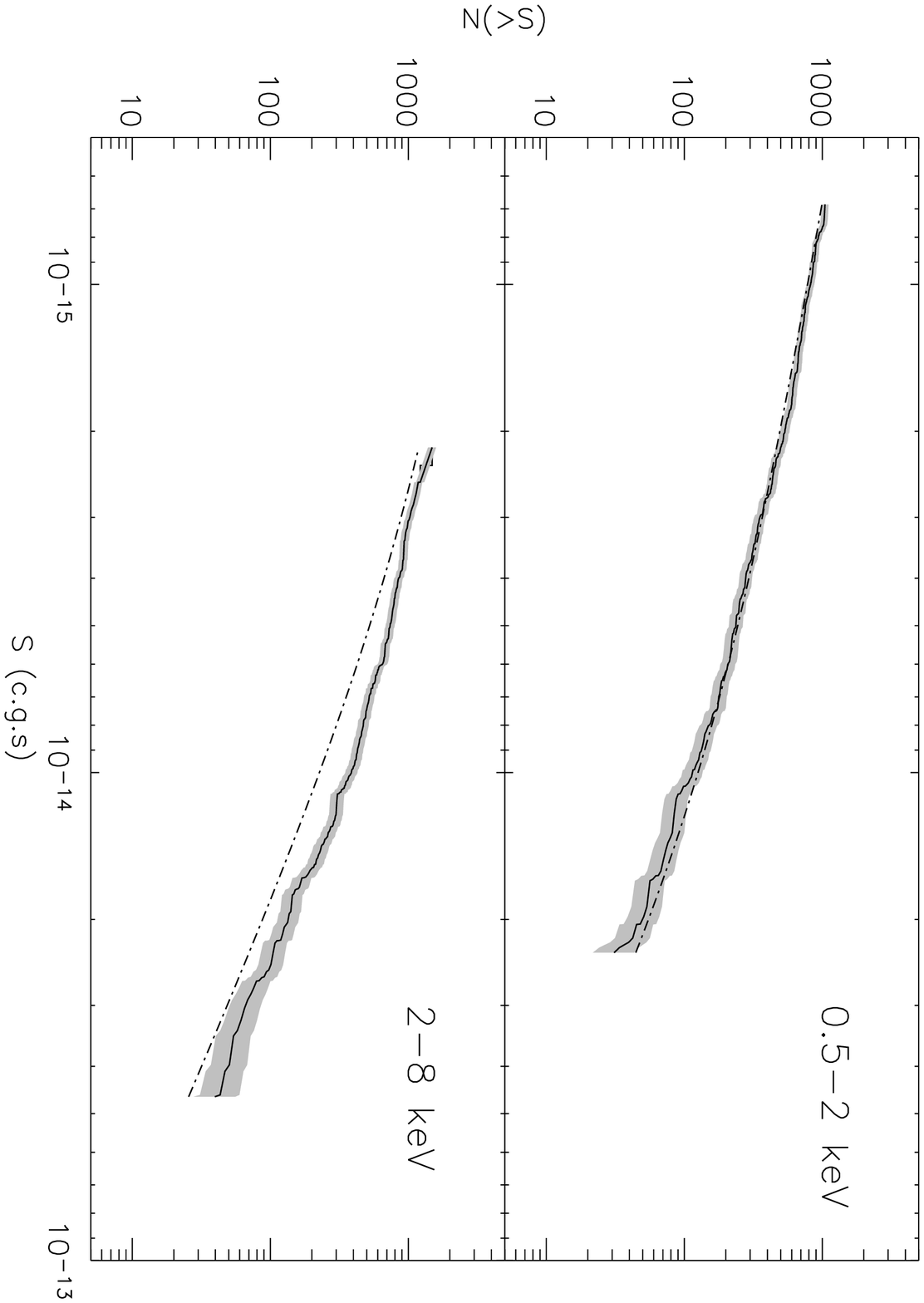] {Cumulative LogN-LogS for the soft and hard bands. 
The $1\sigma$ error is shaded. Dash-dotted line represents the best fit
from Moretti et al.\ (2003). Hard band LogN-LogS from Moretti et al.
is rescaled to that of $2-8$~keV, assuming $\Gamma=1.4$.
\label{fig:LogN-LogS}}

\figcaption[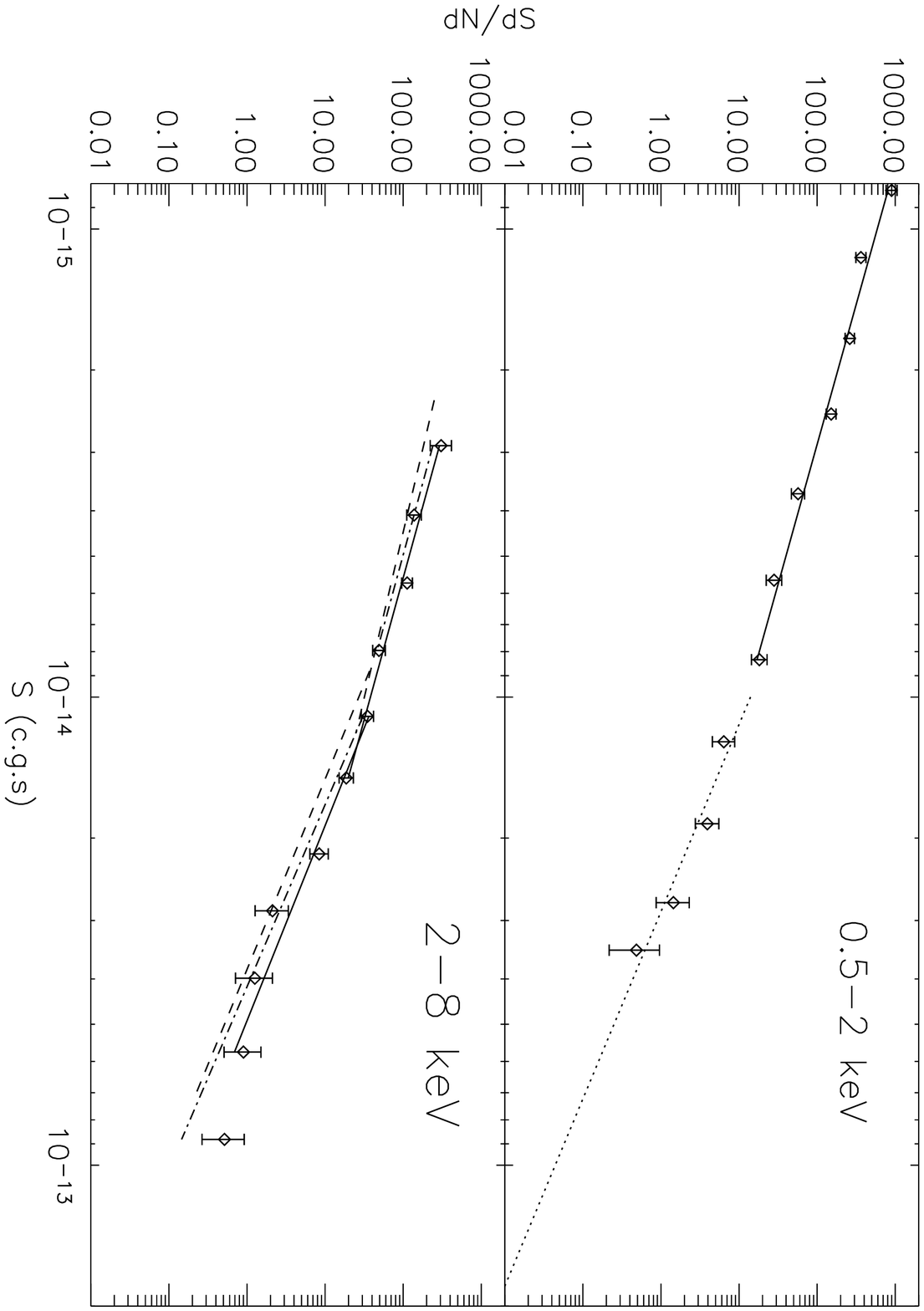] {Differential LogN-LogS for the
soft and hard bands. The unit of dN/dS is number per
$10^{-15}$\ergpcmsqps. Best-fit power laws are shown
as solid lines. Dotted line represents a power law with
a fixed index of $-2.5$. The data do not constrain the
slope at high fluxes well. Dashed line shows the best
fit of the hard band LogN-LogS from the SEXSI survey
(Harrison et al.\ 2003) and the dash-dotted line is the
best fit from Cowie et al. (2002). 
\label{fig:diff_logNlogS }}

\figcaption[fig13.ps] {Hardness ratio vs. full band flux for the 
CLASXS sources. Open circles with arrows represent the upper
or lower limits. Dashed lines with numbers label
the hypothetical spectral indices, assuming
the source spectra are single power laws with only Galactic
absorption. Dotted line represents the typical error
size of the hardness ratio for a source with hardness ratio
of 1.
\label{fig:hardness_vs_flux }}

\figcaption[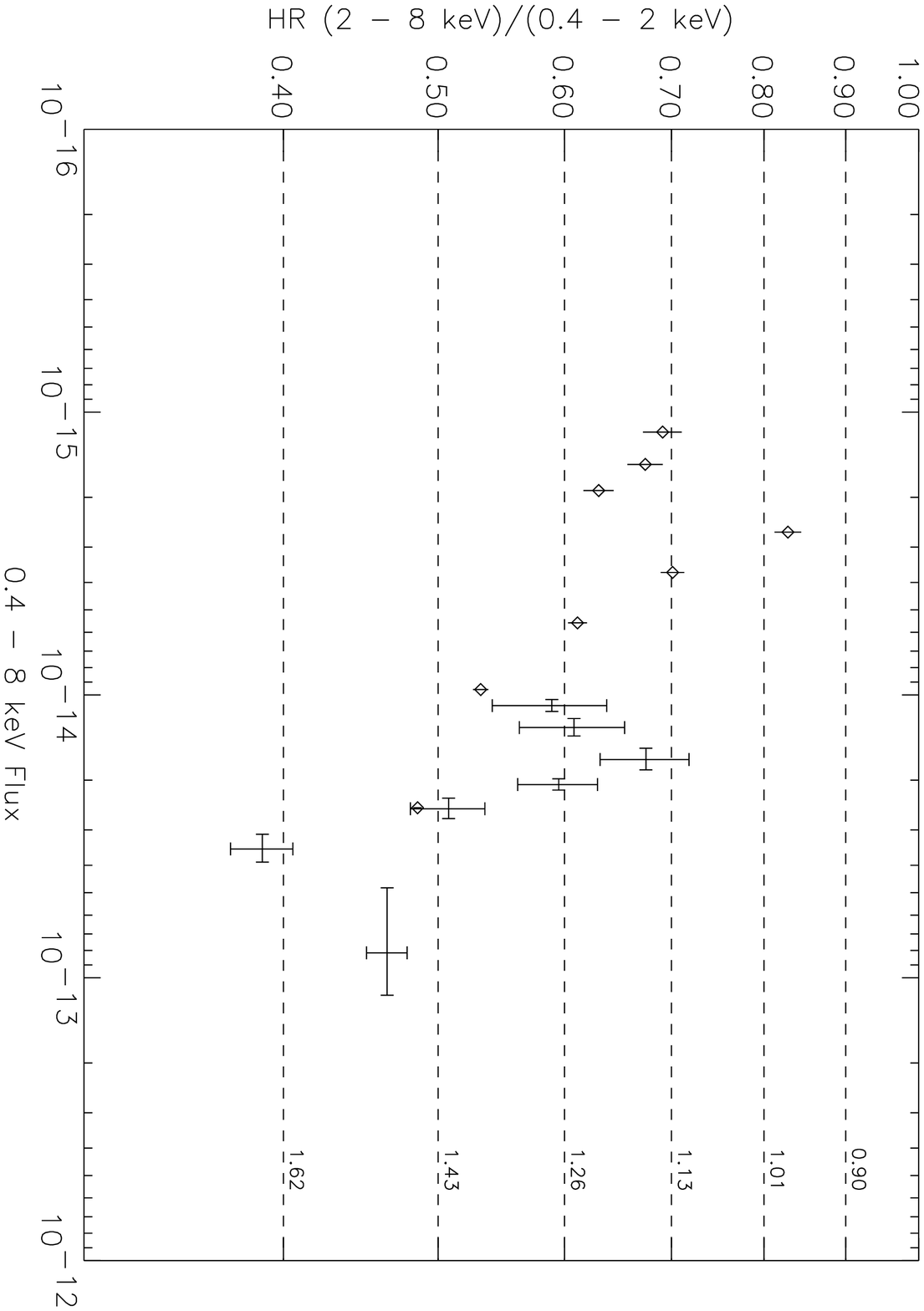] {Hardness ratio of the stacked
sources in different flux bins. Crosses are the
CLASXS sources and diamonds are the combined CDFs sources
(Alexander et al.\ 2003). Sources with fluxes below
$8 \times 10^{-15}$~\ergpcmsqps in the CLASXS catalog and
$1 \times 10^{-15}$~\ergpcmsqps in the CDFs catalogs are
not included to avoid incompleteness. Dashed lines
are as in Figure 10.
\label{fig:stacked }}

\figcaption[fig15.ps] {Light curves of the sources
detected to be variable. The fluxes are normalized
to the mean of all the observations. Numbers in the plots
are the  source numbers in the catalog. (a) Soft band;
(b) hard band; (c) full band.
\label{fig:lightcurve }}

\figcaption[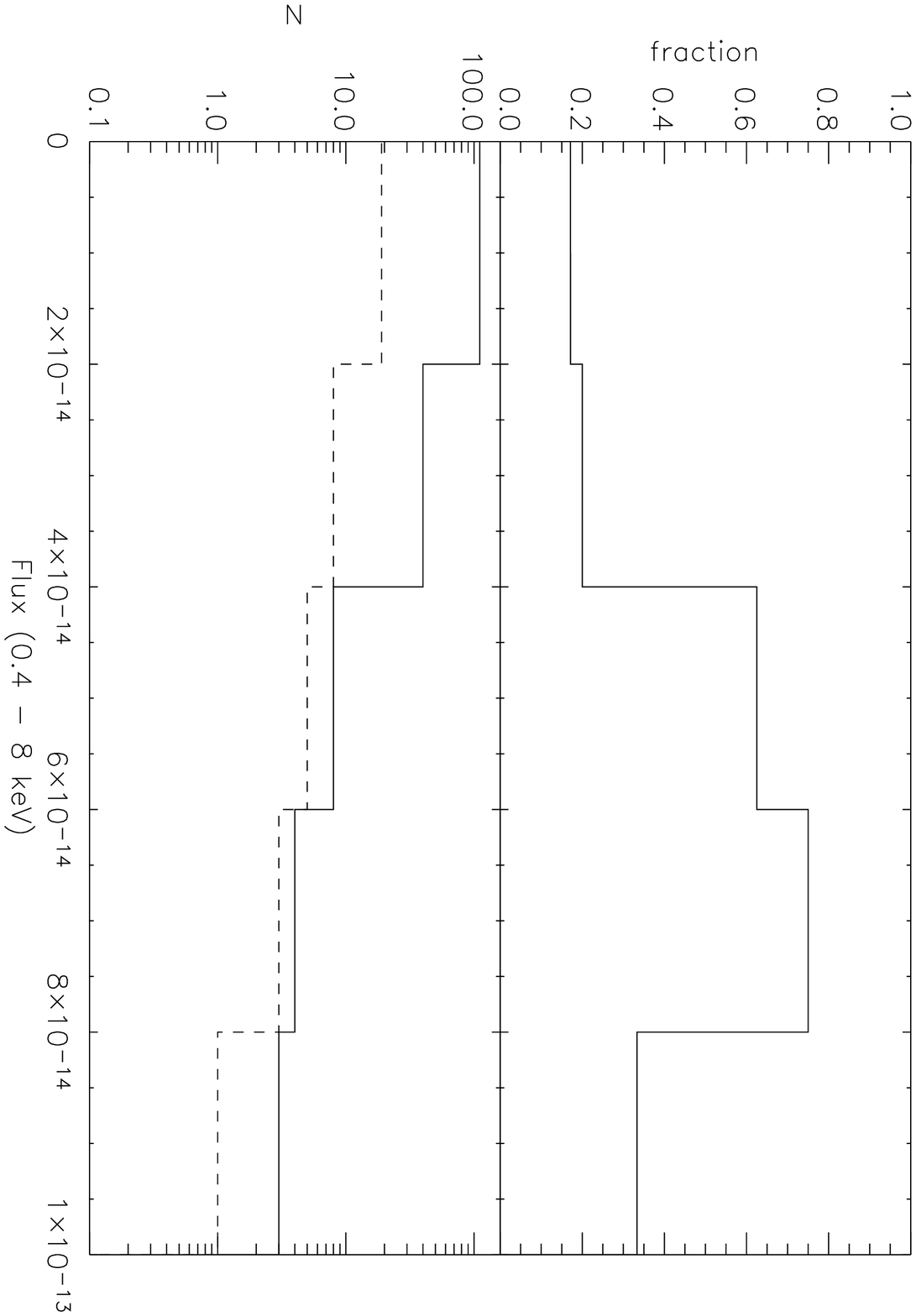] {{\em (Upper panel)} Fraction of sources
that are variable in different flux bins. {\em (Lower panel)} 
Number of variable sources {\em (dashed histogram)}
and total number of sources tested for variability
{\em (solid histogram)} in the same flux bins as in the upper panel.
\label{fig:fraction_var }}

\figcaption[fig17.ps] {Excess variability for the variable sources
in each energy band vs. the flux of that band. (a) Soft band;
(b) hard band; (c) full band.
\label{fig:excess variability }}

\figcaption[fig18.ps] {Spectral variability vs.
full band fluxes for all the variable sources. The fluxes are in
units of $10^{-14}$\ergpcmsqps. Numbers on top of each plot are the
source numbers in the catalog.
\label{fig:spectral_var }}

\figcaption[fig19.ps] {Adaptively smoothed X-ray images of the extended
sources superposed on R band images.(a) Sources 1 and 2; (b) Source 3; 
(c) Source 4. 
\label{fig:cluster_images }}

\figcaption[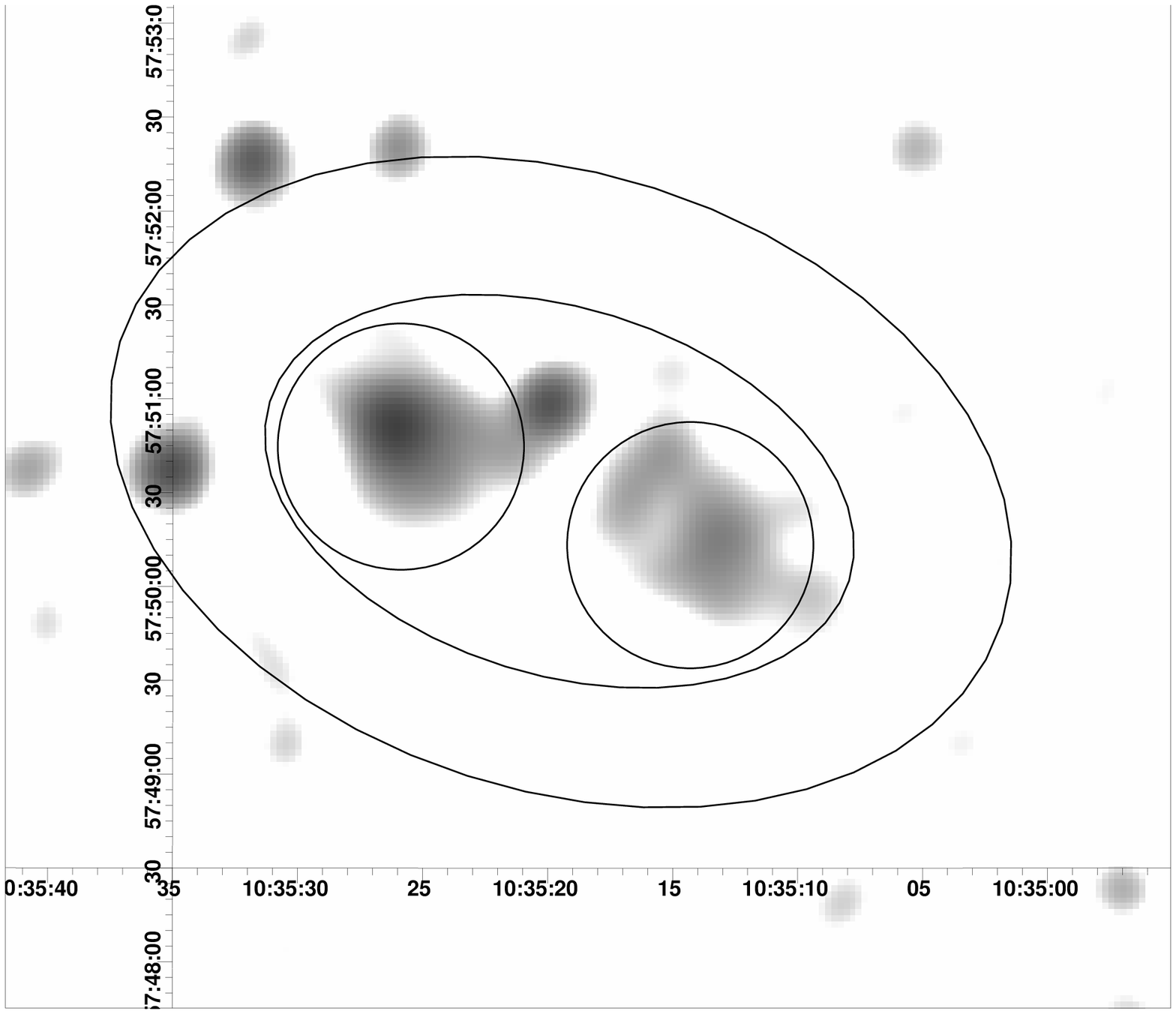] {  Regions for spectral extraction of Sources 1 and 2
on the Gaussian smoothed gray scale map of the clusters.
The Gaussian kernal size is 6\arcsec. Source regions
are shown as circles. Elliptical annulus region is for the background
extraction.
\label{fig: cluster1and2_extract}}

\figcaption[fig21.ps] { X-ray spectra and best-fit MEKAL models of 
Source 1 {\em (dash-dotted line)} and Source 2 {\em (solid line)}.
\label{fig:spec_1and2 }}

\figcaption[fig22.ps] { Combined probability contour of the 
temperature of Sources 1 and 2.
Contour lines are 1, 2, and 3$\sigma$ confidence levels. Cross is the 
best-fit temperature.
Solid line represents the equality of the temperature of the two clusters.
\label{fig:TvsTplot }}

\figcaption[fig23.ps] {X-ray spectrum of Source 3 and the best-fit 
MEKAL model.
\label{fig:spec_cluster3 }}

\figcaption[fig24.ps] {Curve-of-growths for the extended sources (Sources
1 -- 4 shown in pannels (a) --(d)) normalized
to the best-fit background. Dotted line shows the best fit of
an integrated 2 dimensional Gaussian.
\label{fig: curve_of_growth}}

\figcaption[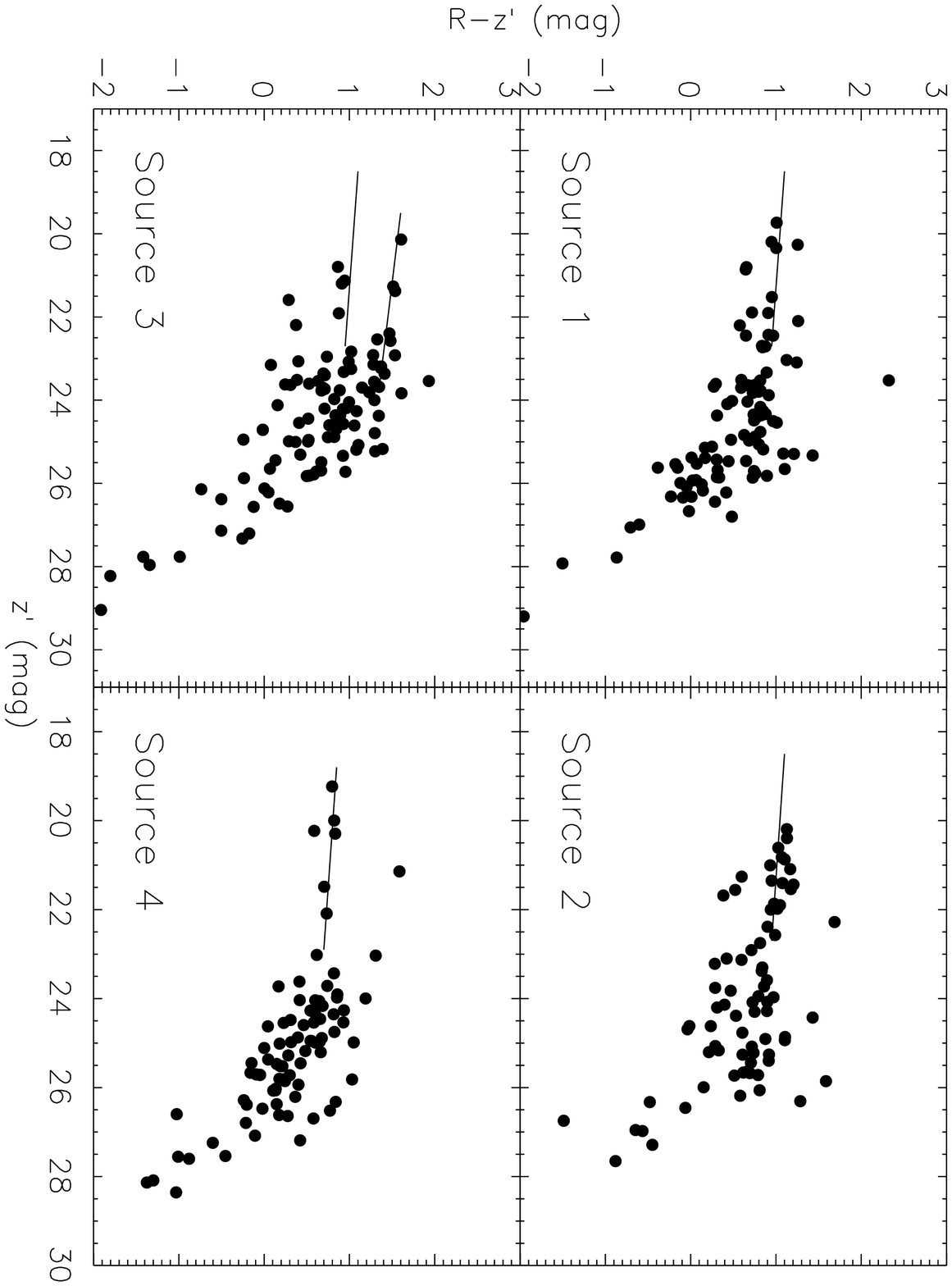] {Color-magnitude plot for the galaxies within 
0.5\arcmin~of the X-ray center. Solid lines show the model red sequence from
Yee \& Gladders (2001) at the redshifts that best match the observations in
Source 1 ($z=0.5$), 2 ($z=0.5$), and 4 ($z=0.45$).  In the plot for 
Source 4, the red sequences for $z=0.5$ (lower solid line) and $z=1.0$ 
(upper solid line) are shown.
\label{fig:red_sequence }}

\figcaption[fig26.ps] {$R$ band image of the gravitational lensing 
arc found associated with Source 3.}

\plotone{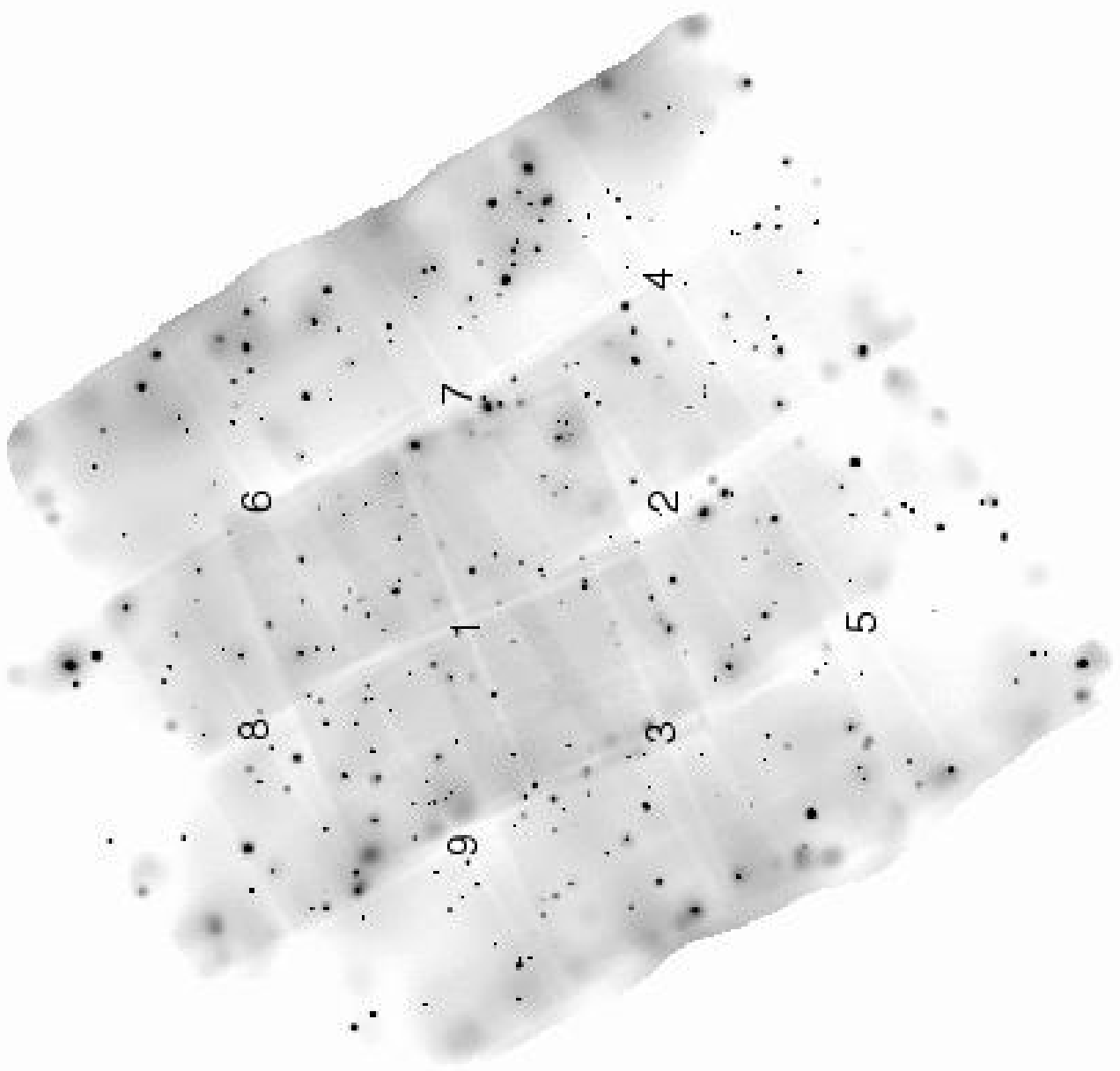}
\plotone{fig2.ps}
\plotone{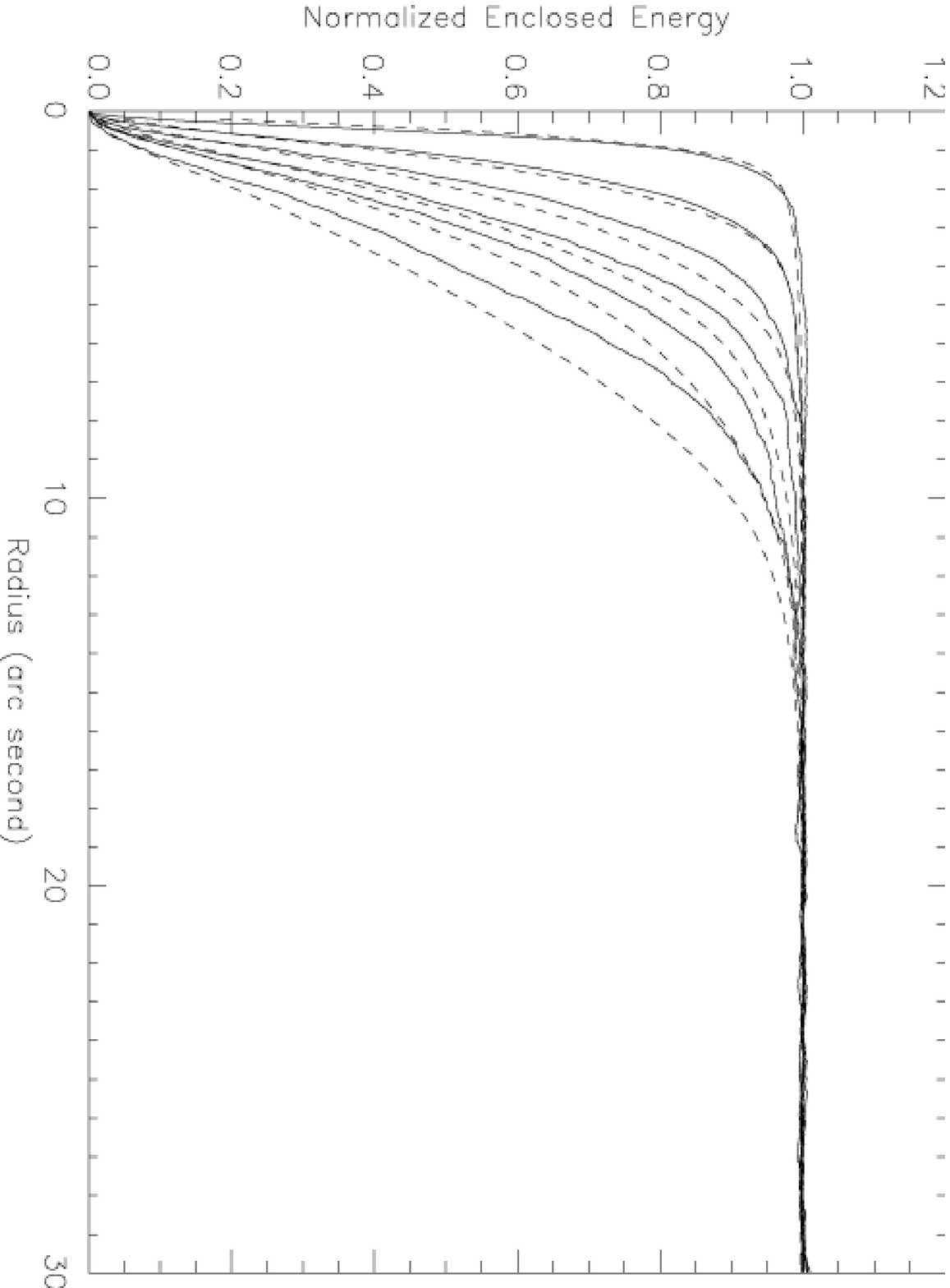}
\plotone{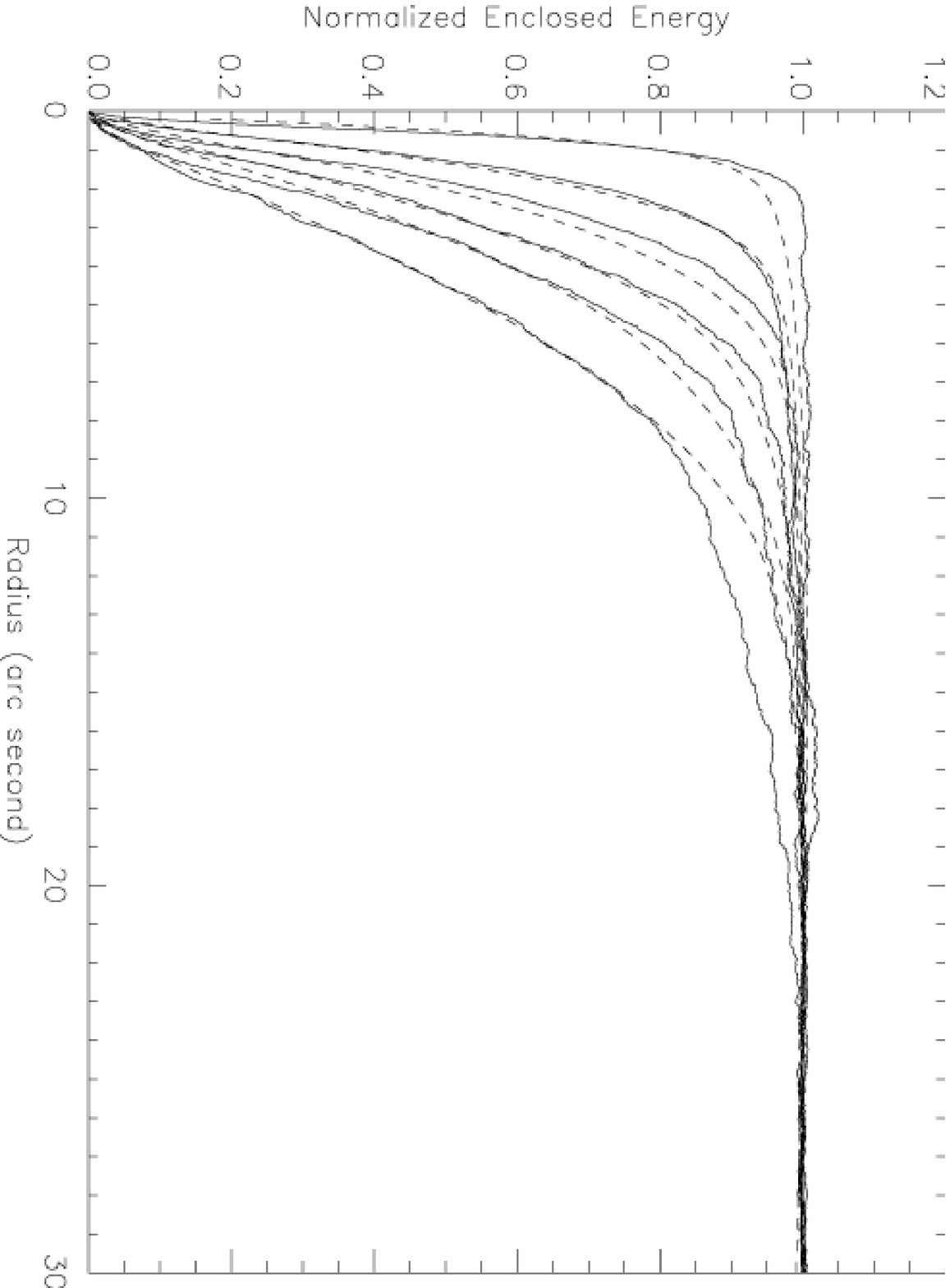}
\plotone{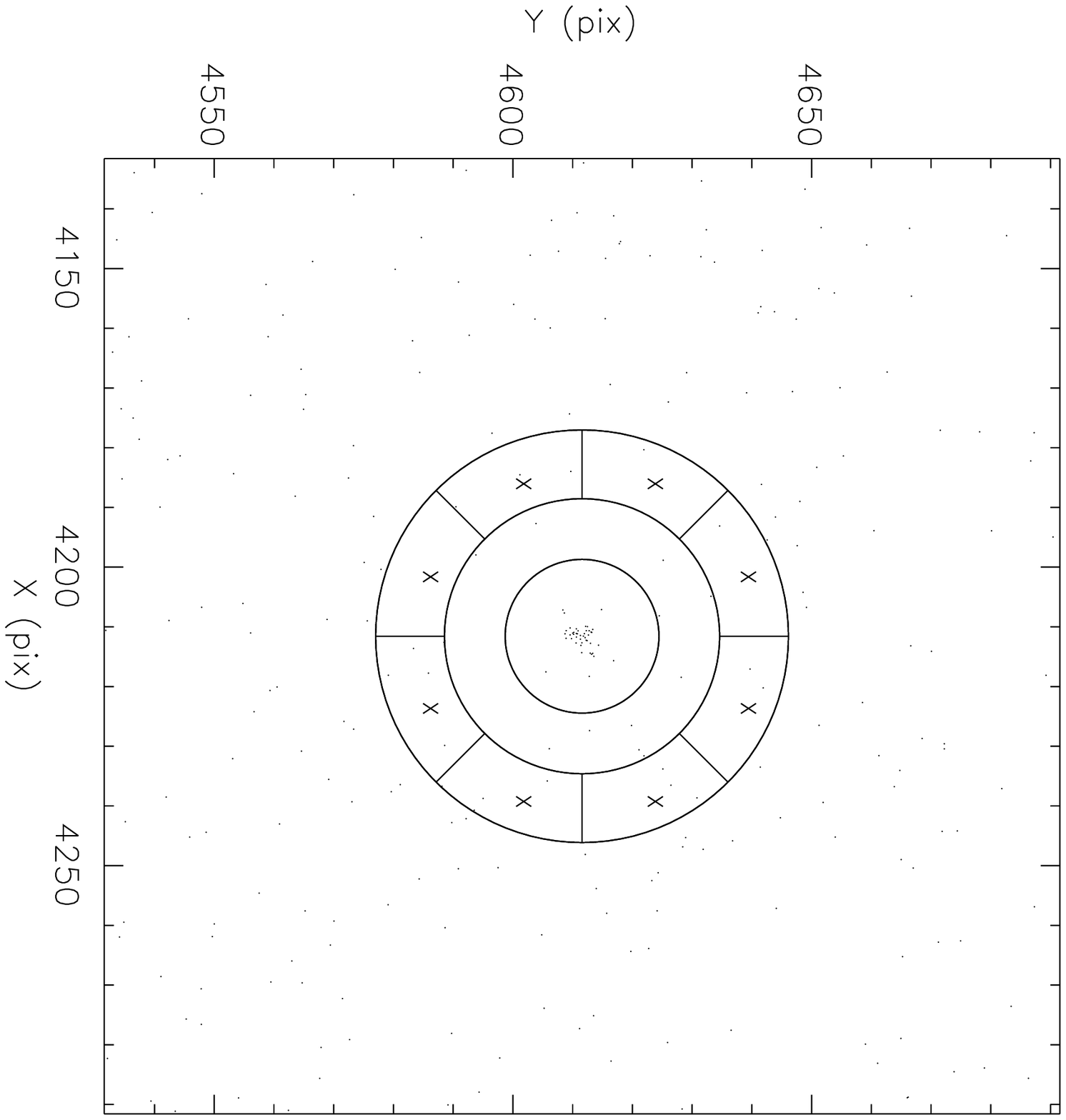}
\plotone{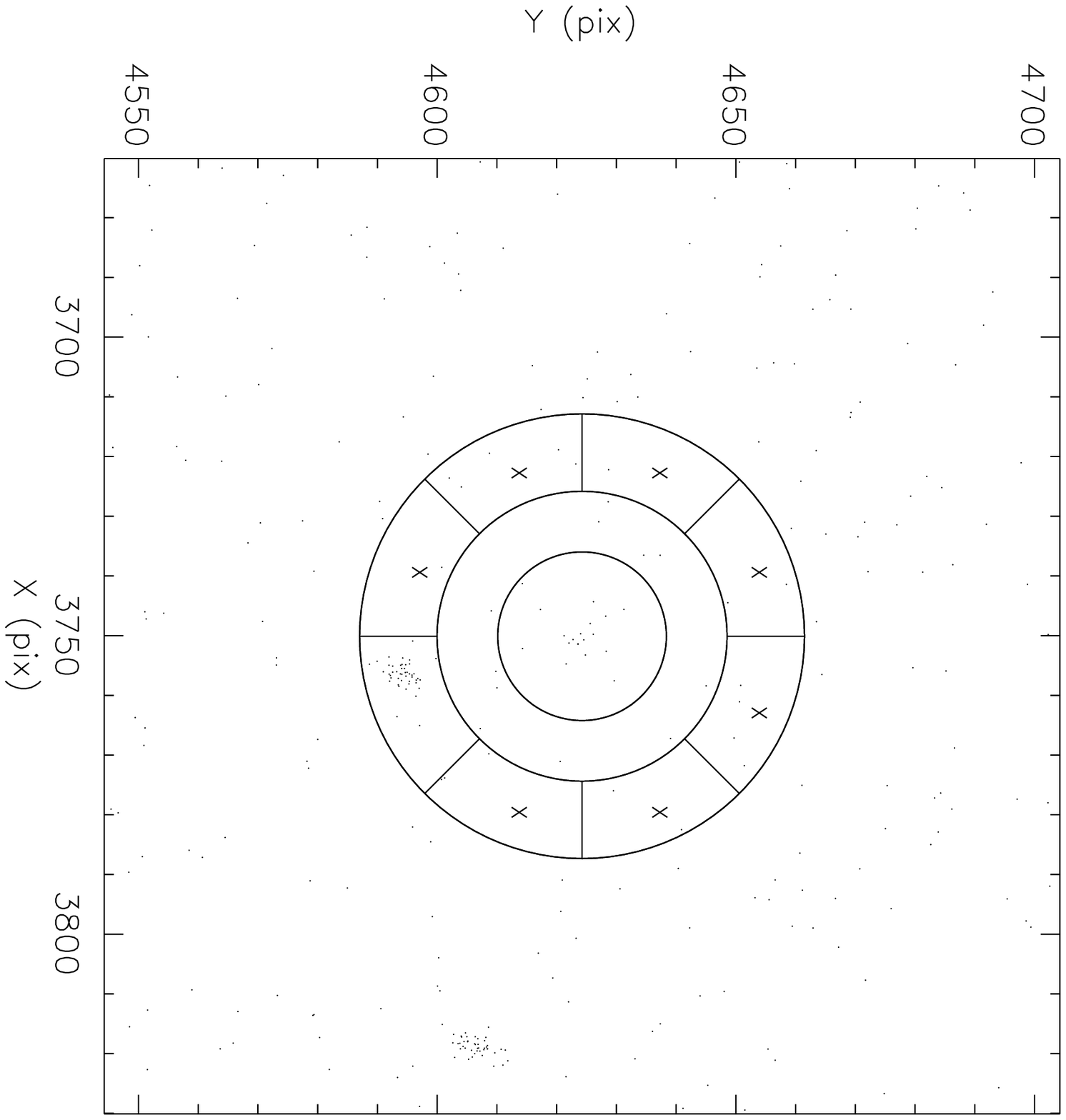}
\plotone{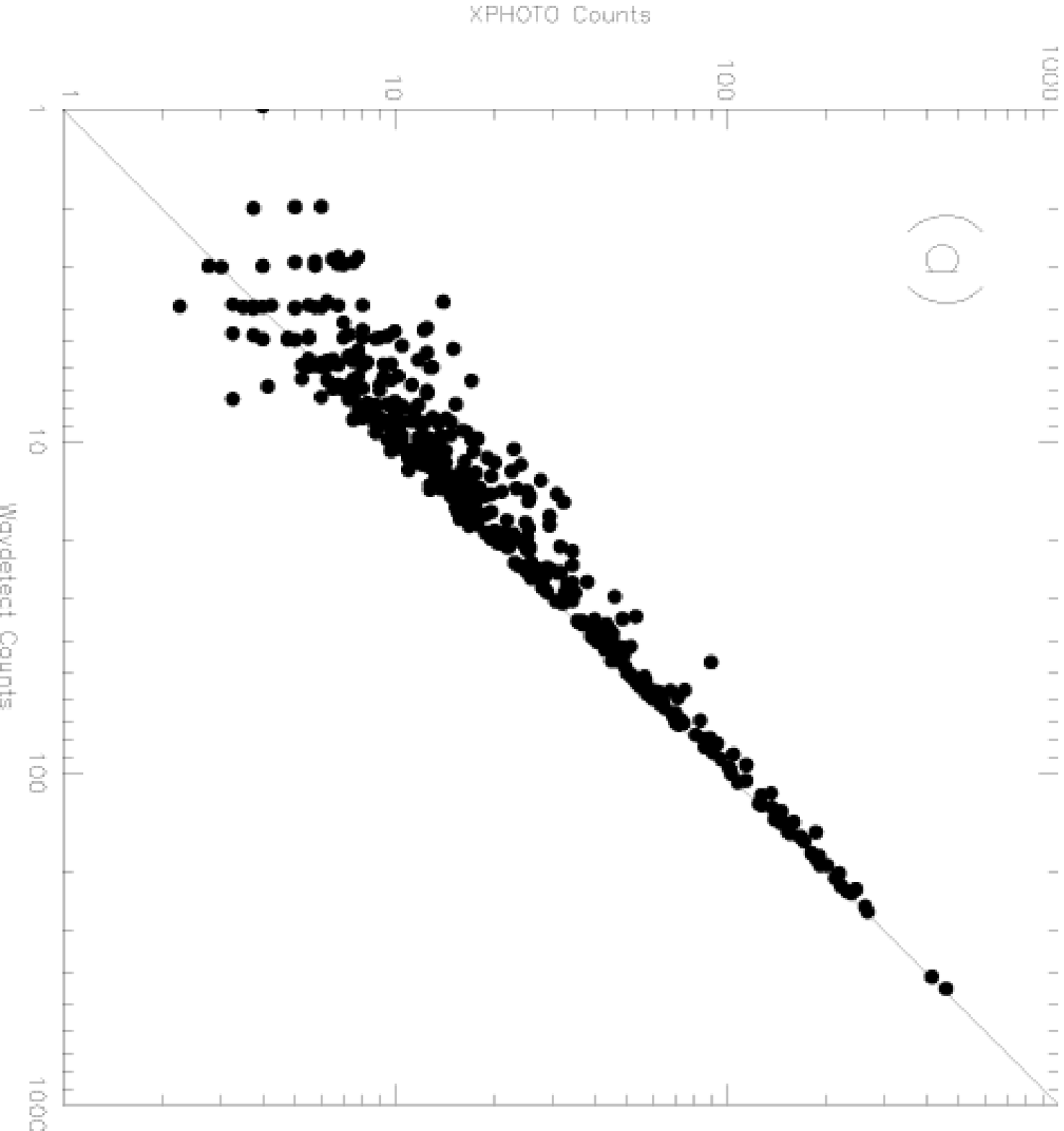}
\plotone{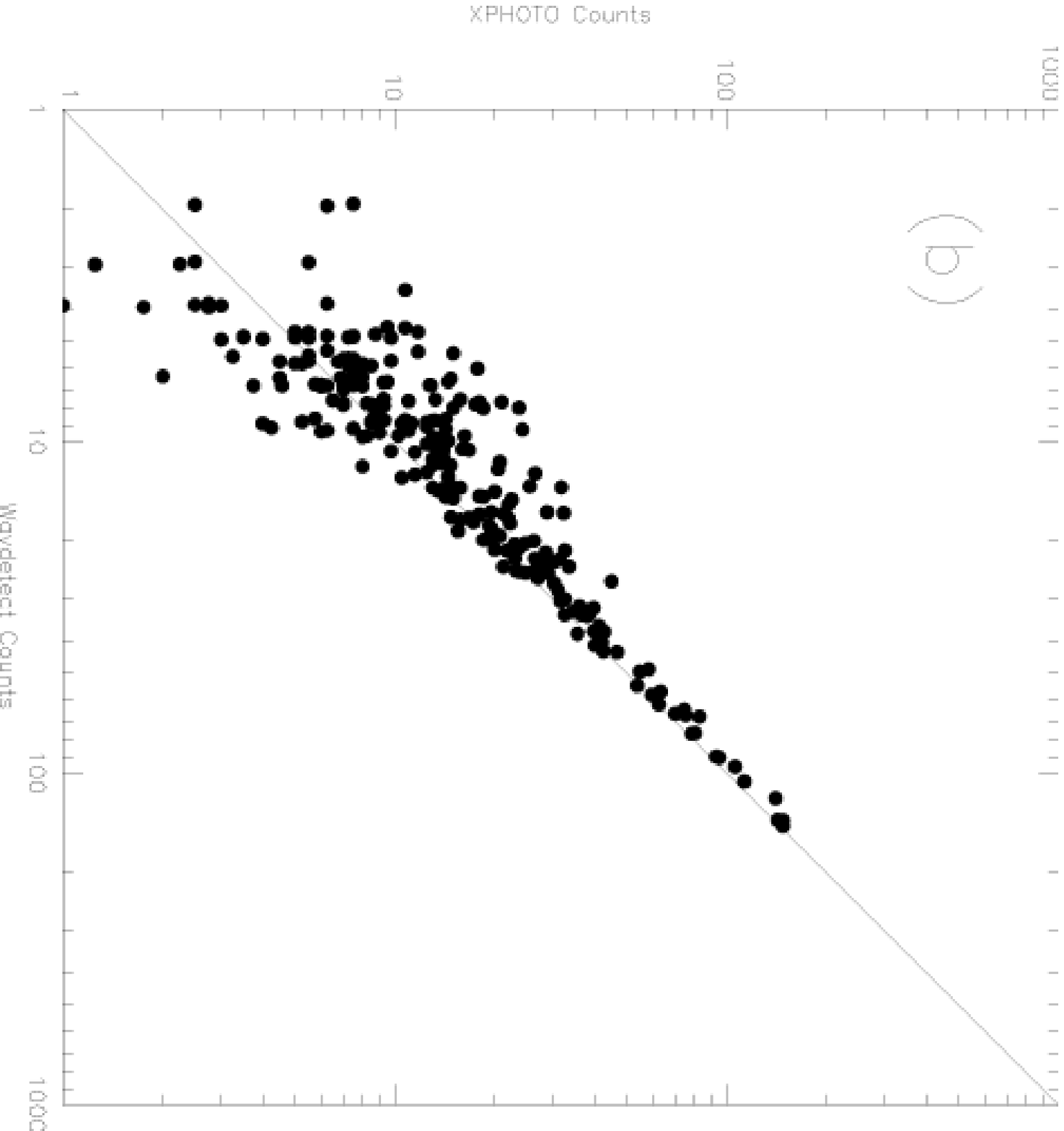}
\plotone{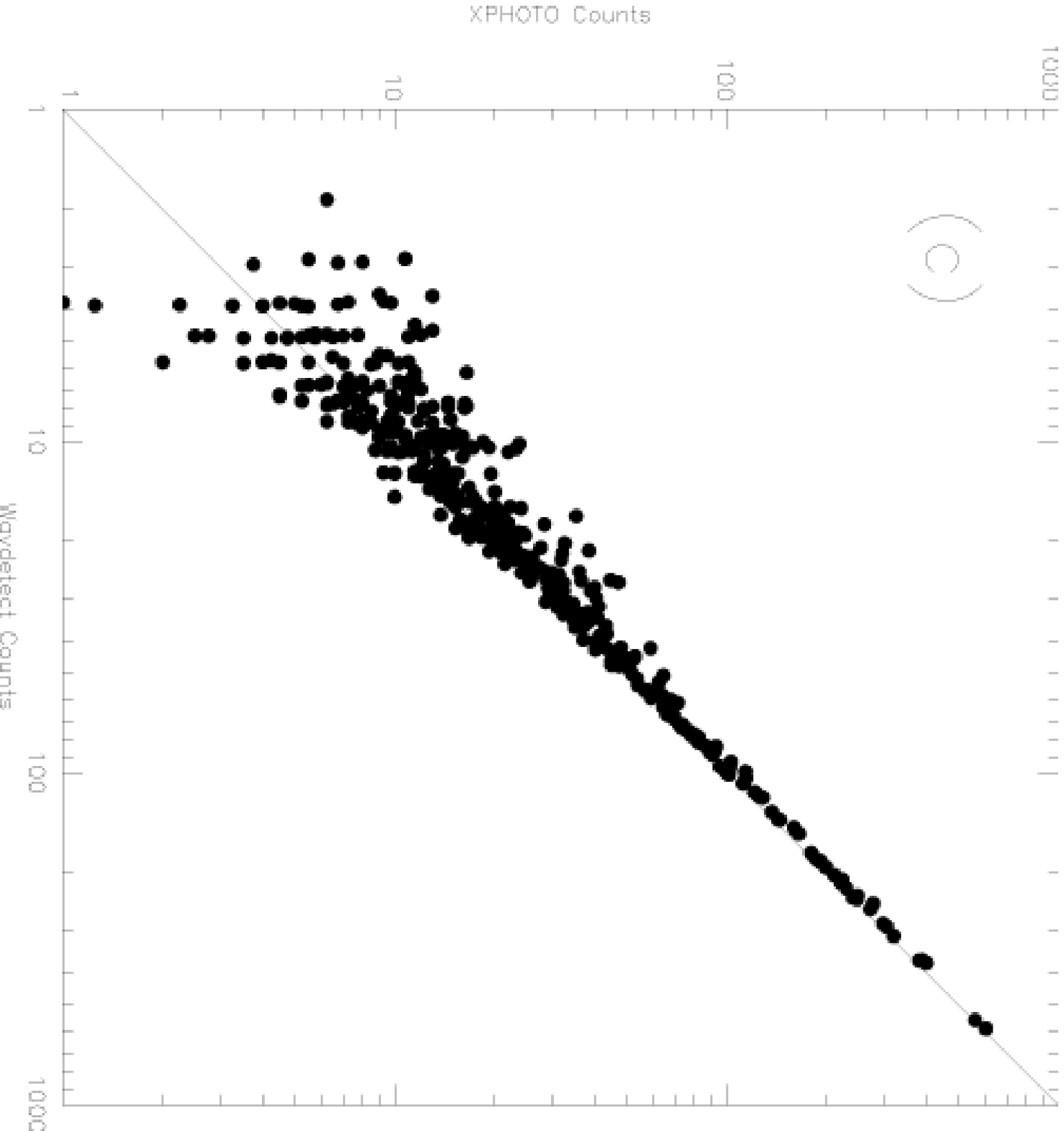}
\plotone{fig6.ps}
\plotone{fig7.ps}
\plotone{fig8.ps}
\plotone{fig9.ps}
\plotone{fig10.ps}
\plotone{fig11.ps}
\plotone{fig12.ps}
\epsscale{0.95}
\plotone{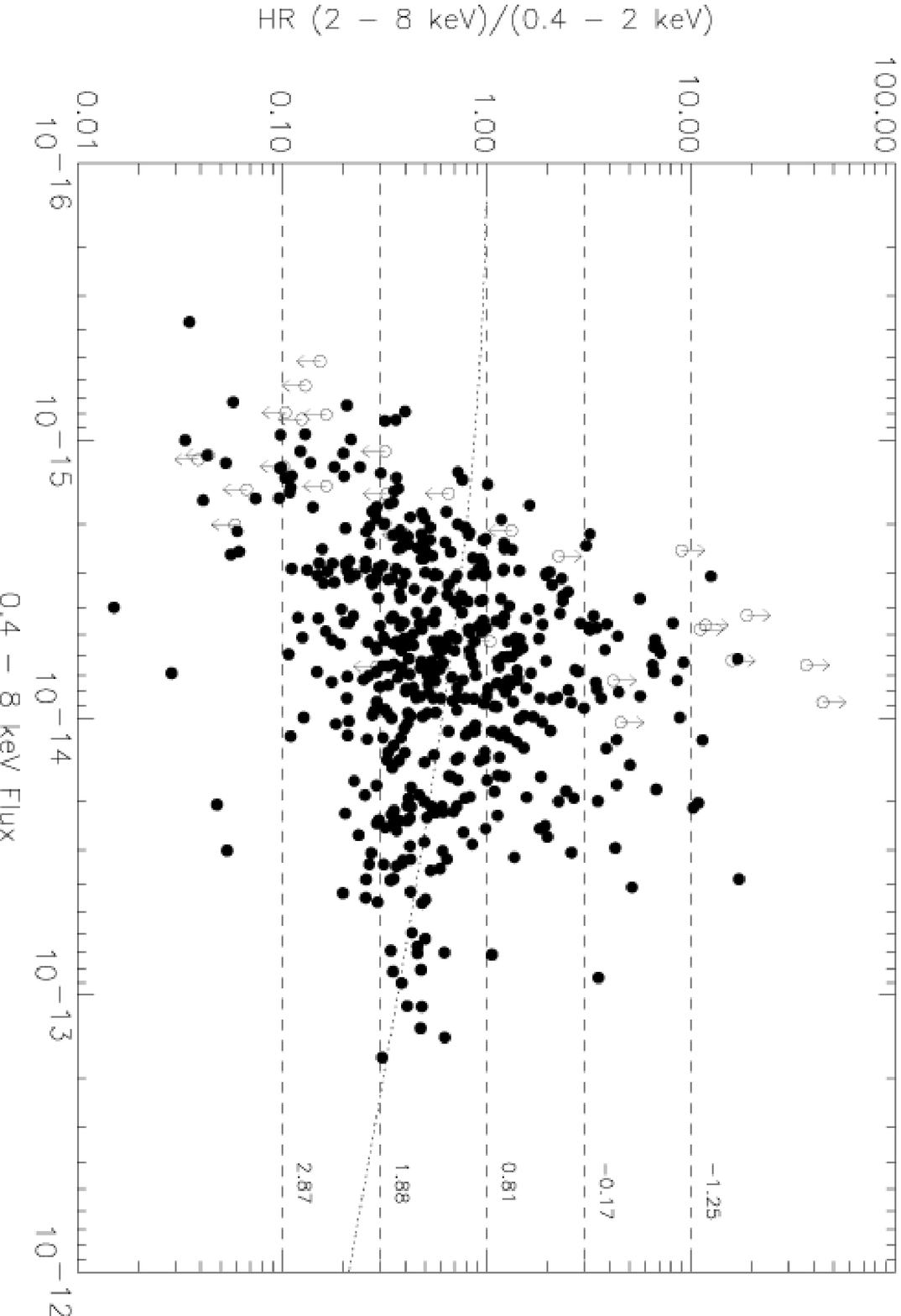}
\epsscale{1.0}
\plotone{fig14.ps}
\plotone{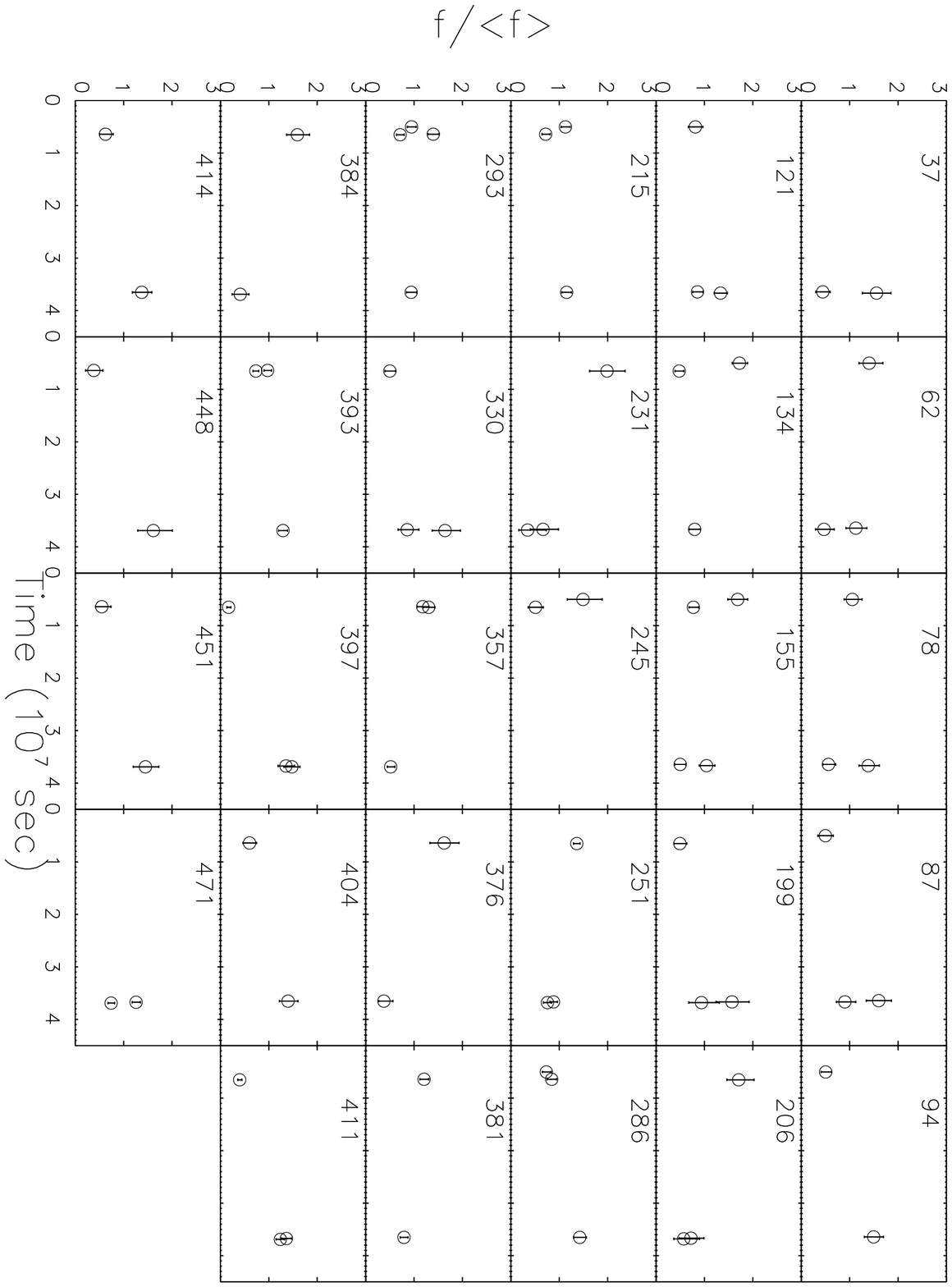}
\plotone{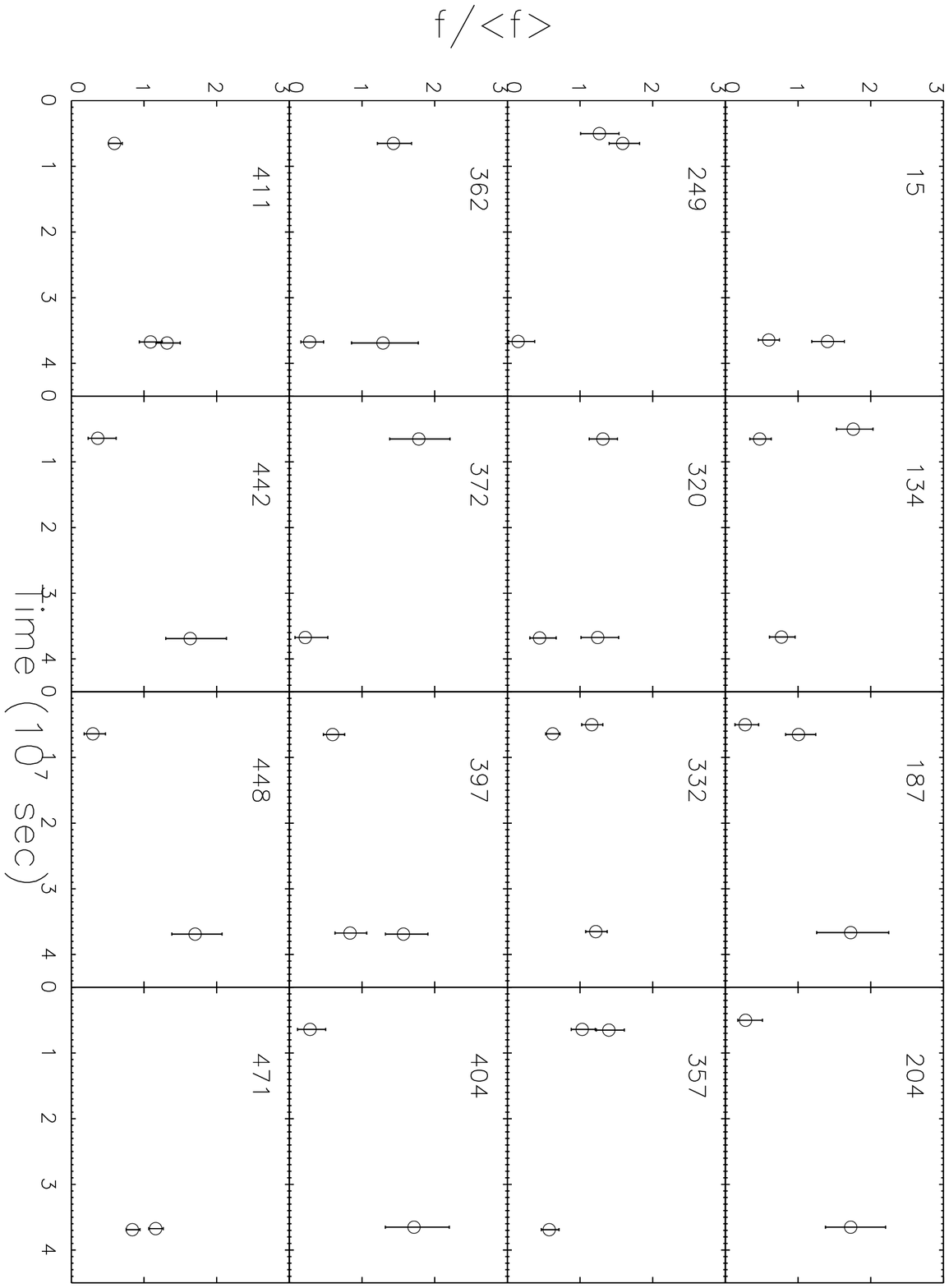}
\plotone{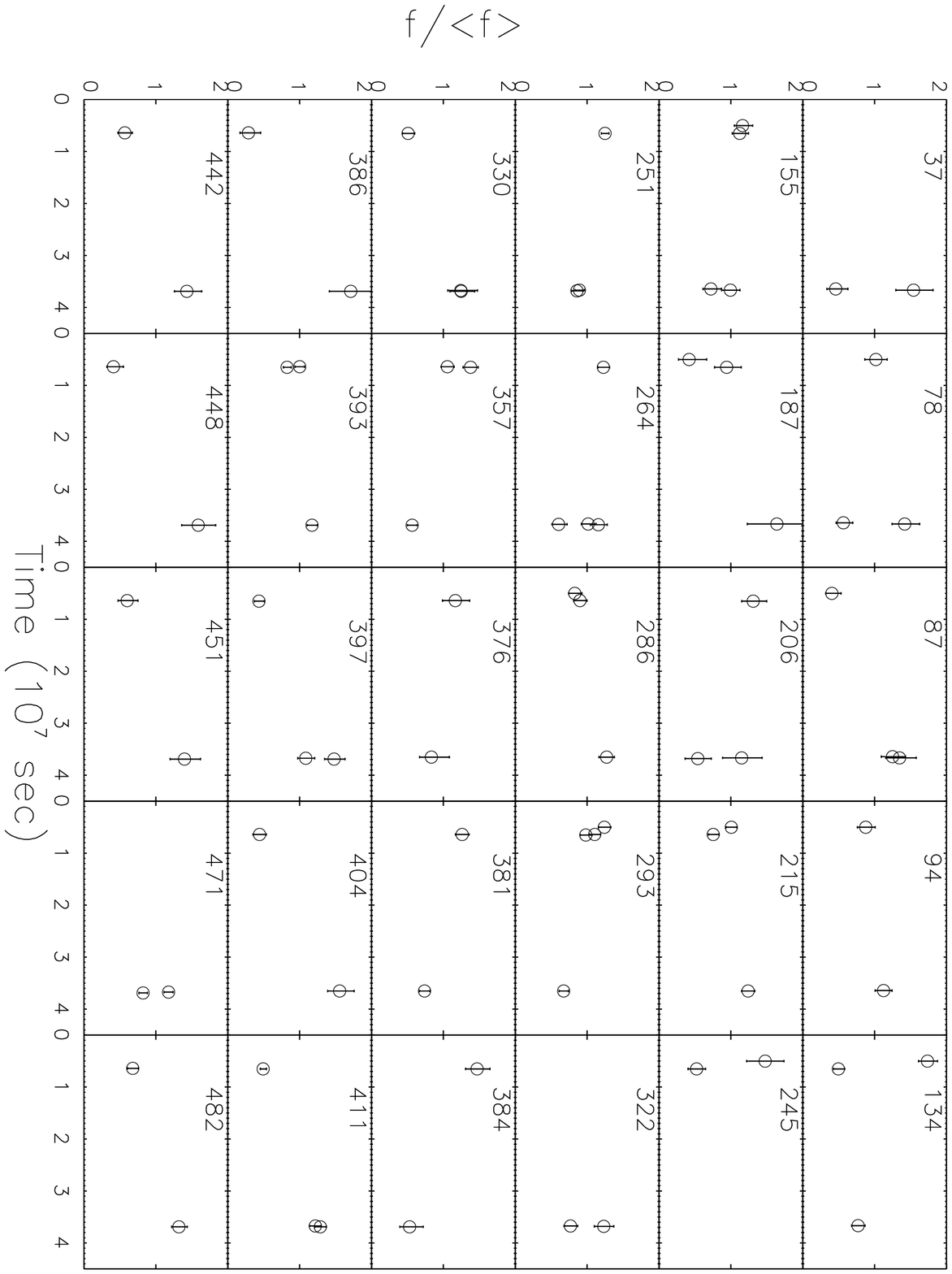}
\plotone{fig16.ps}
\plotone{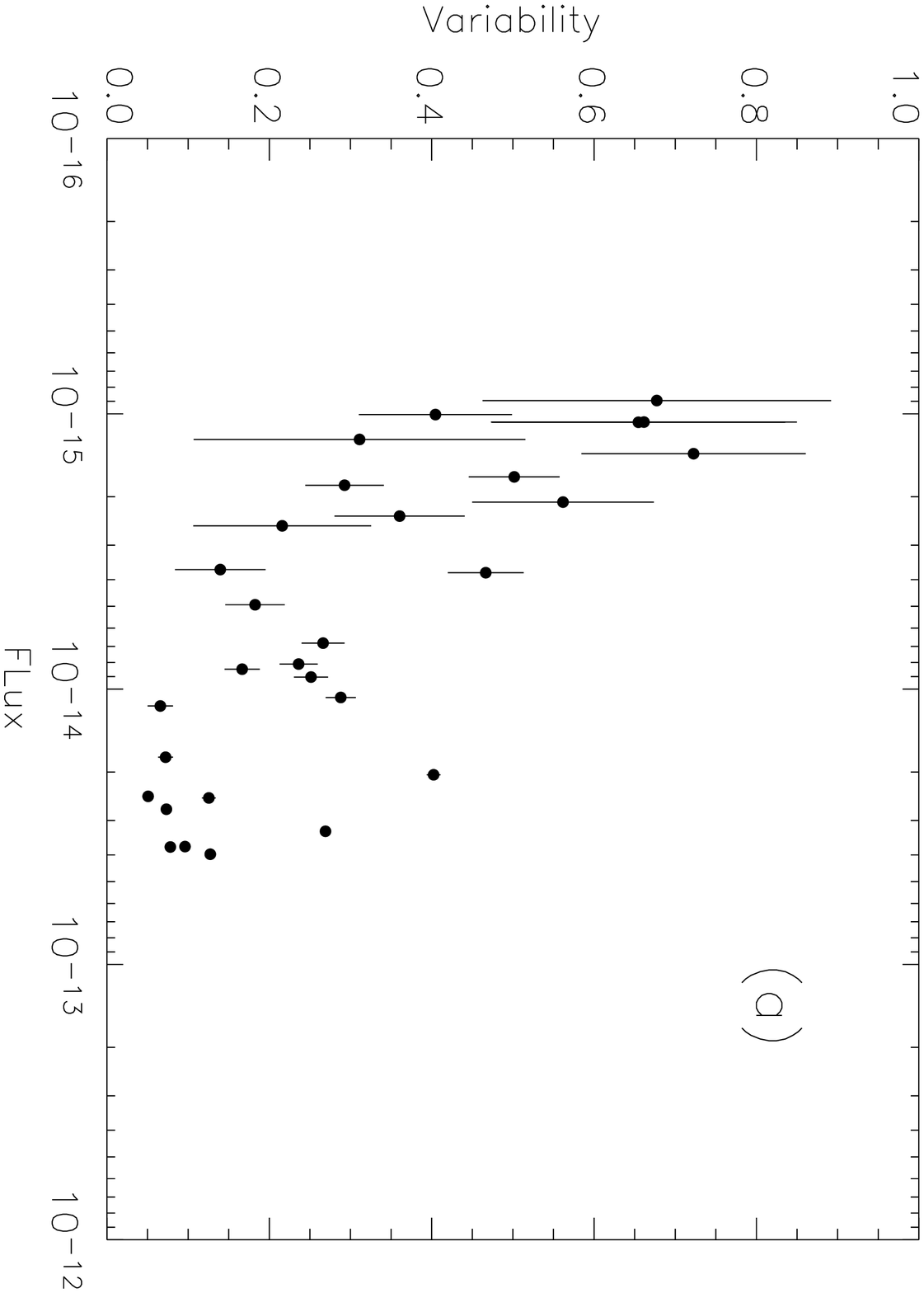}
\plotone{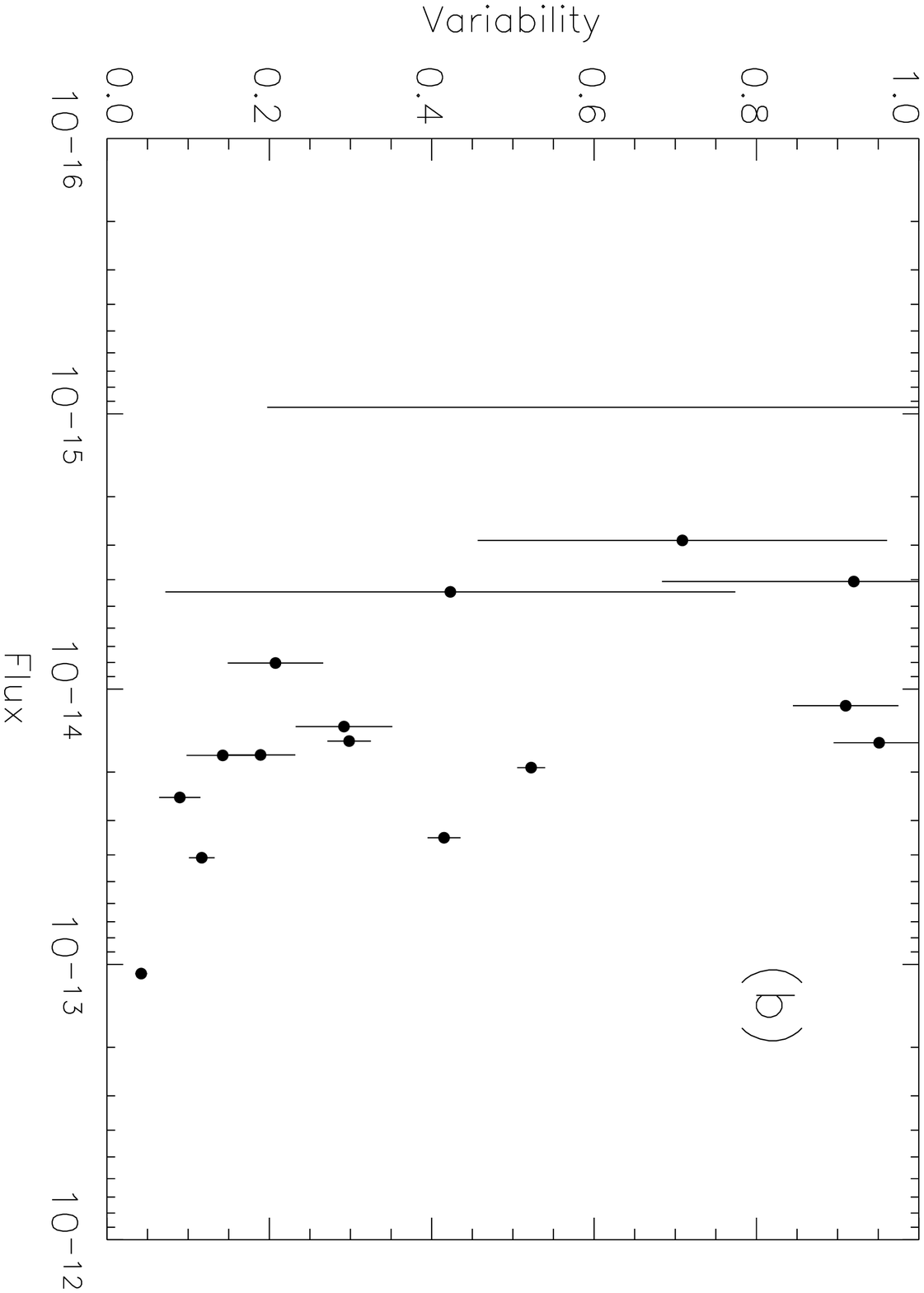}
\plotone{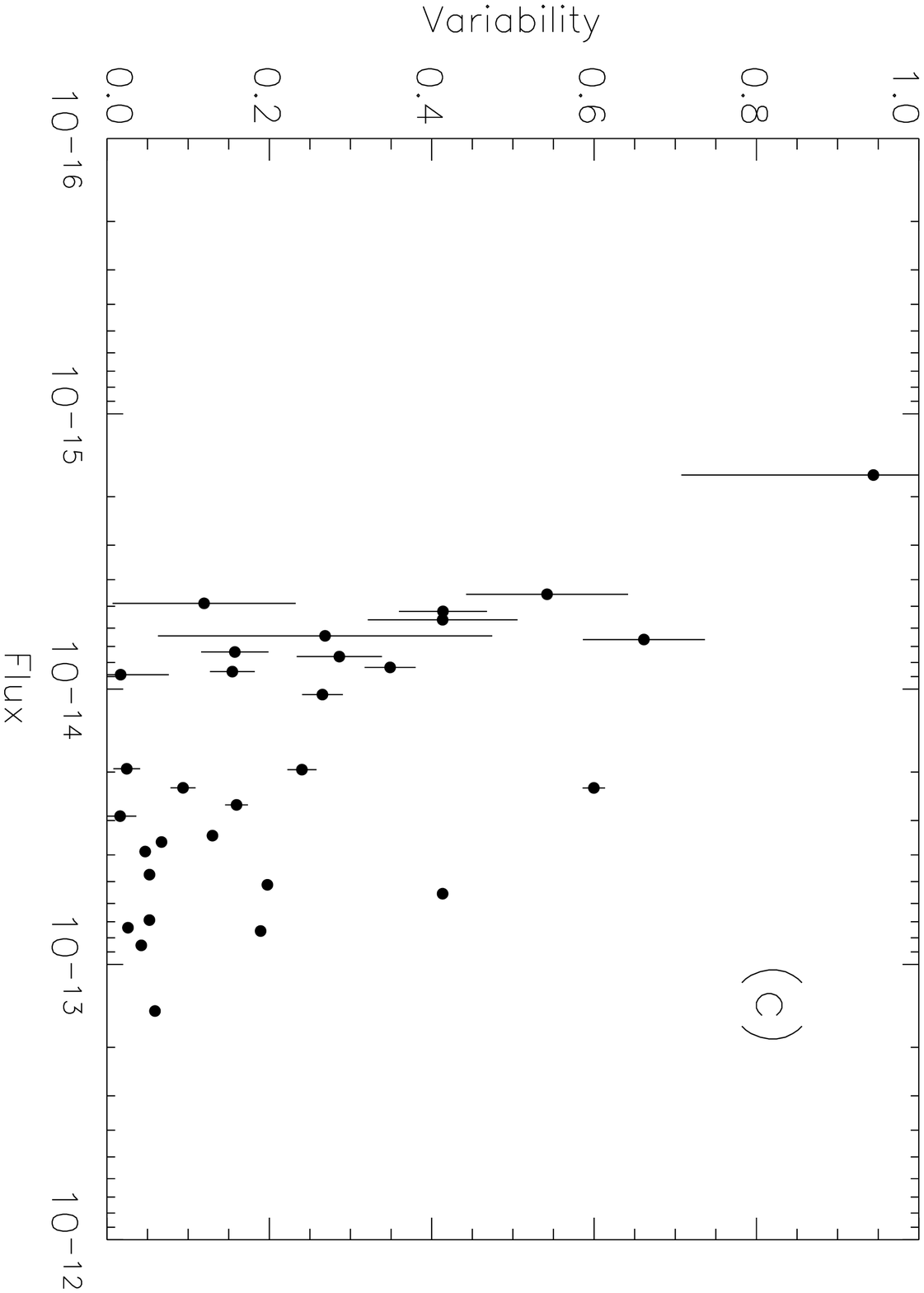}
\plotone{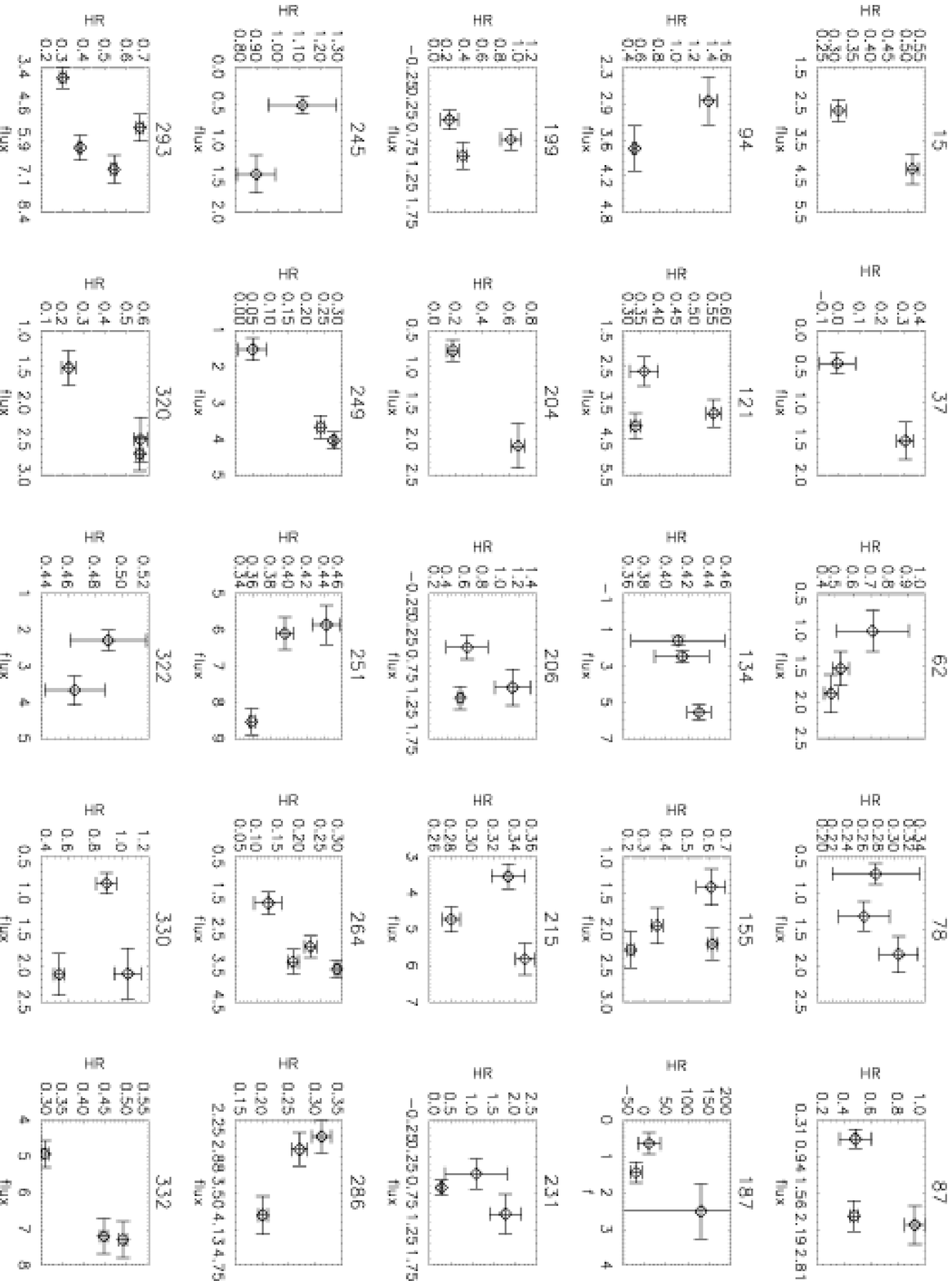}
\epsscale{0.8}
\plotone{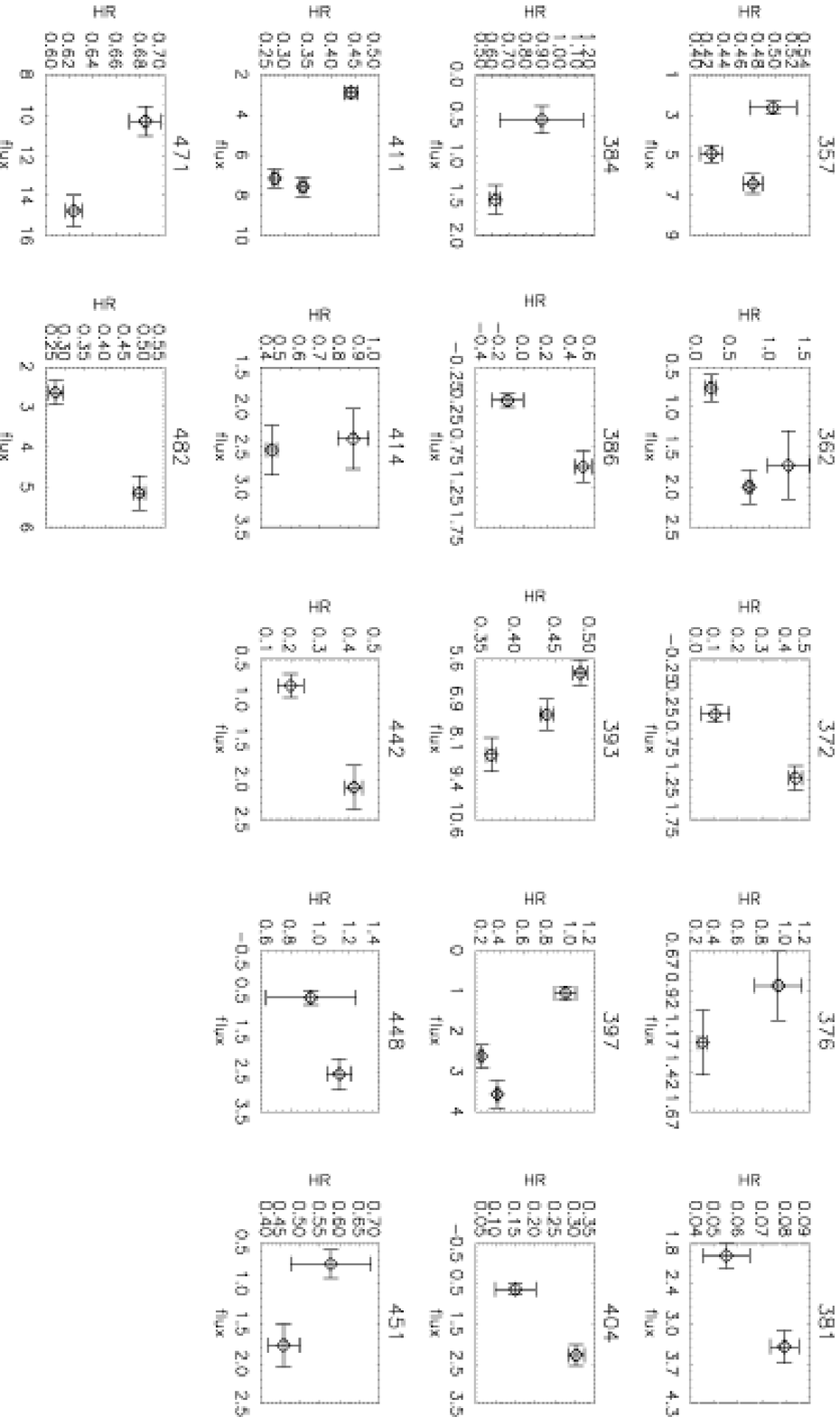}
\epsscale{1.0}
\plotone{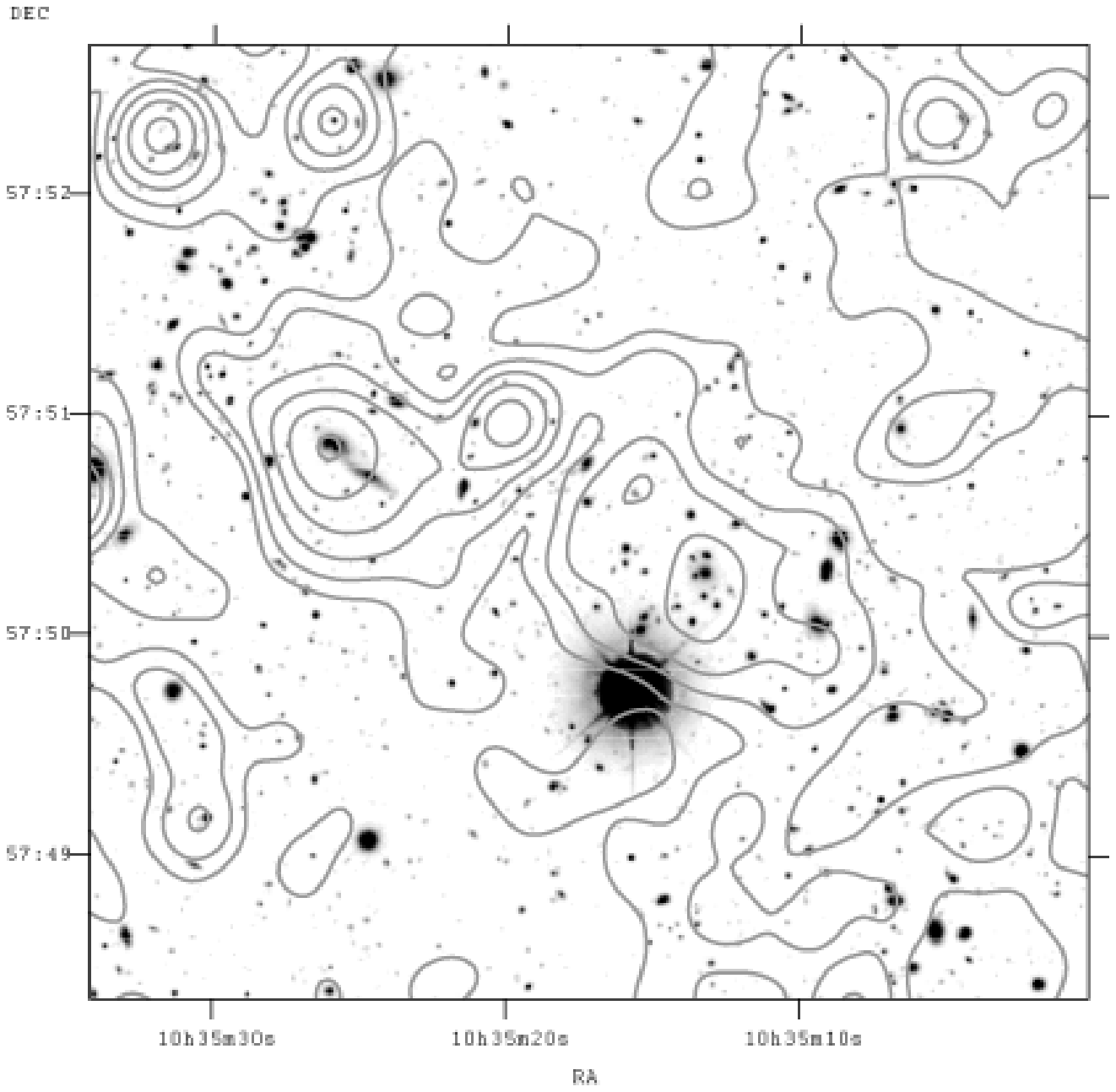}
\plotone{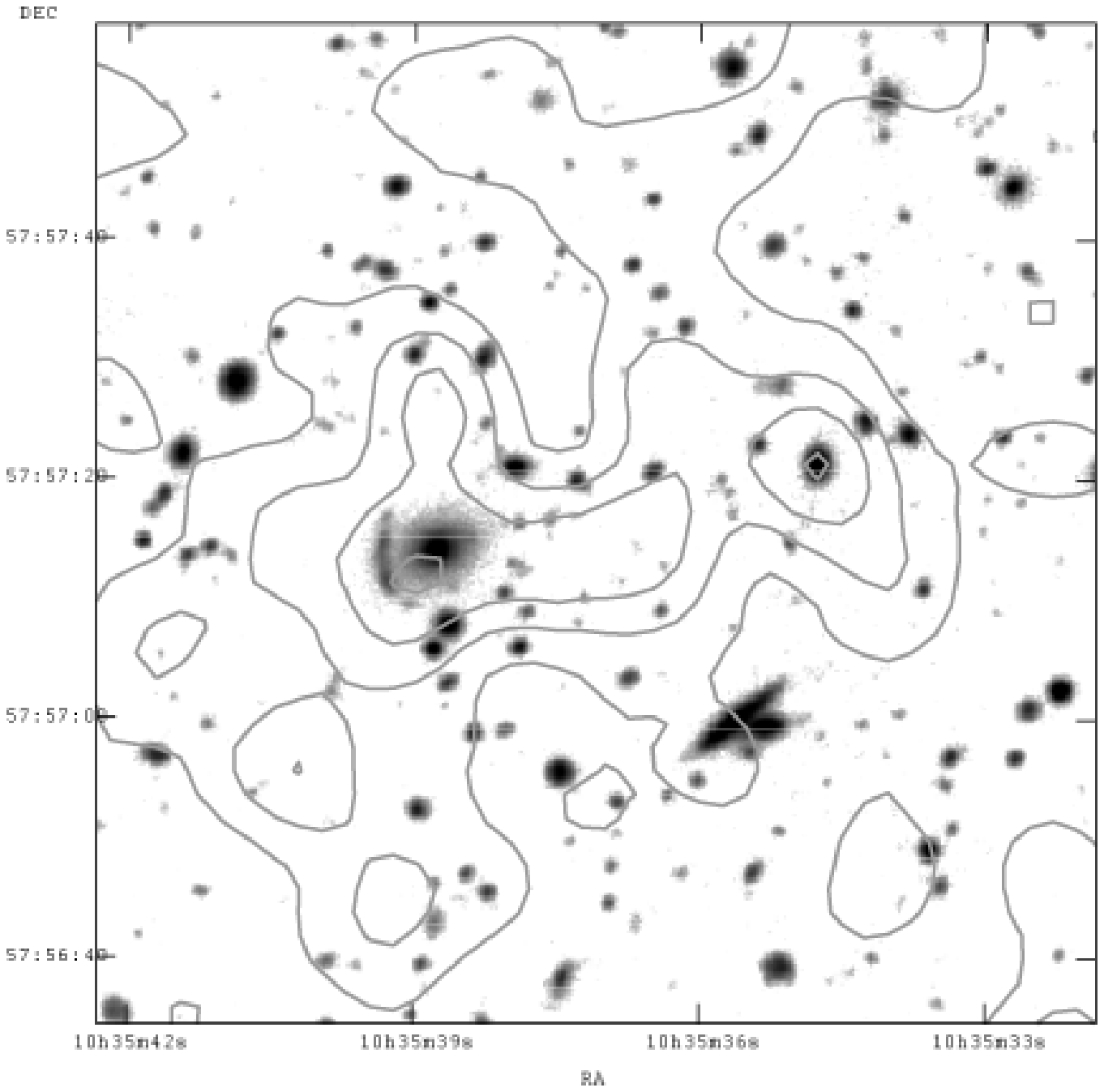}
\plotone{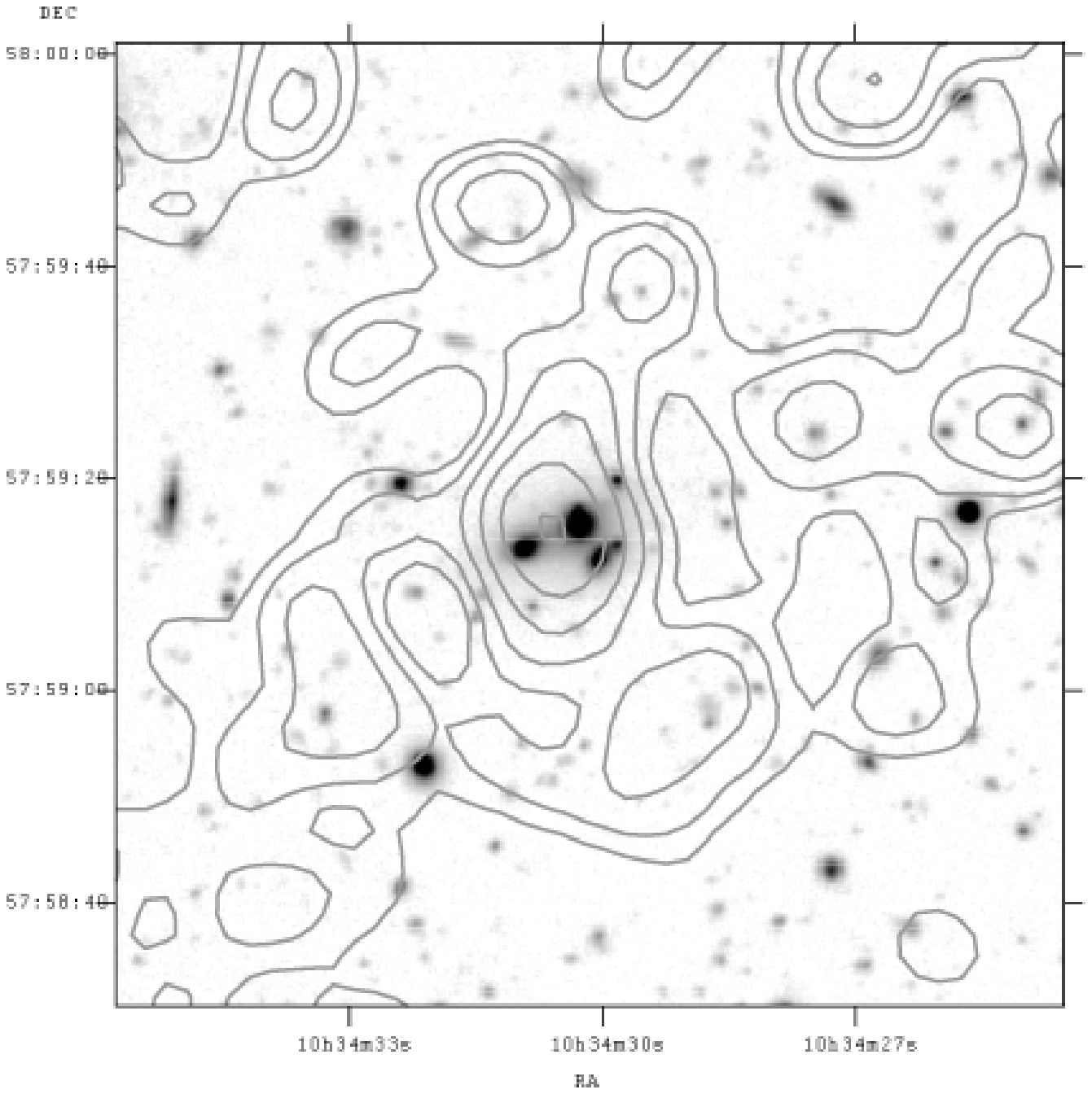}
\plotone{fig20.ps}
\epsscale{0.9}
\plotone{fig21.ps}
\epsscale{1.0}
\plotone{fig22.ps}
\epsscale{0.9}
\plotone{fig23.ps}
\epsscale{1.0}
\plotone{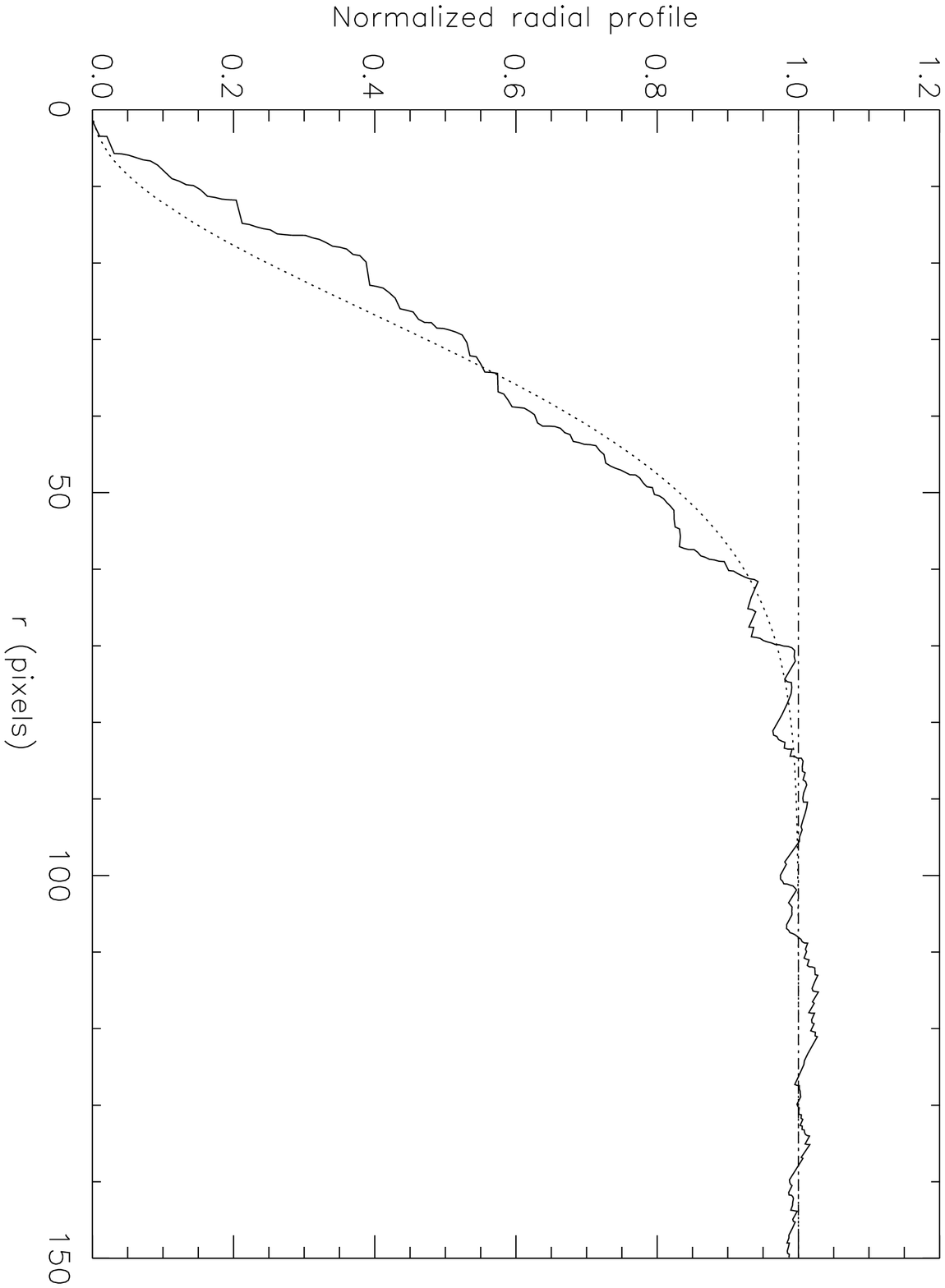}
\plotone{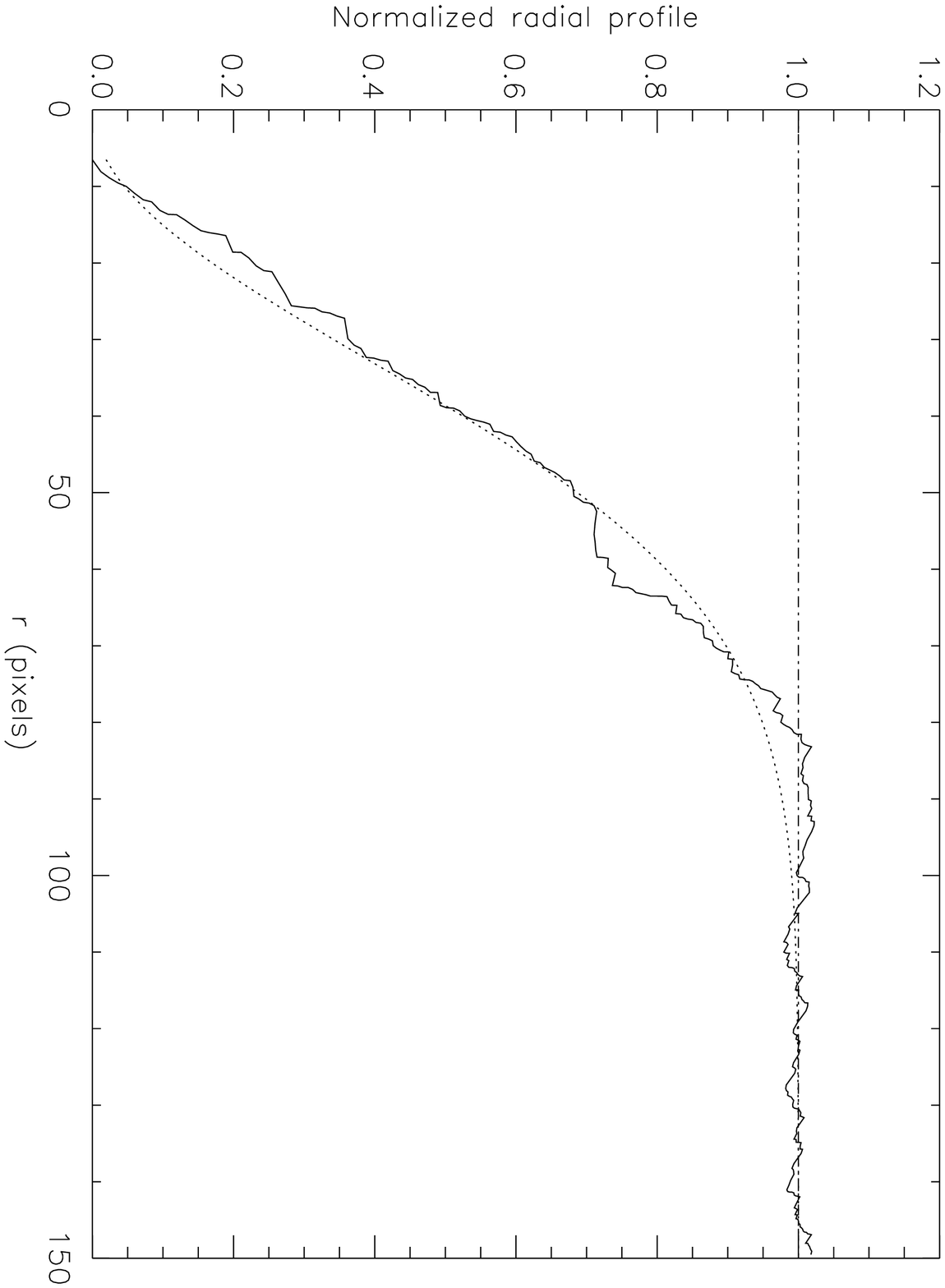}
\plotone{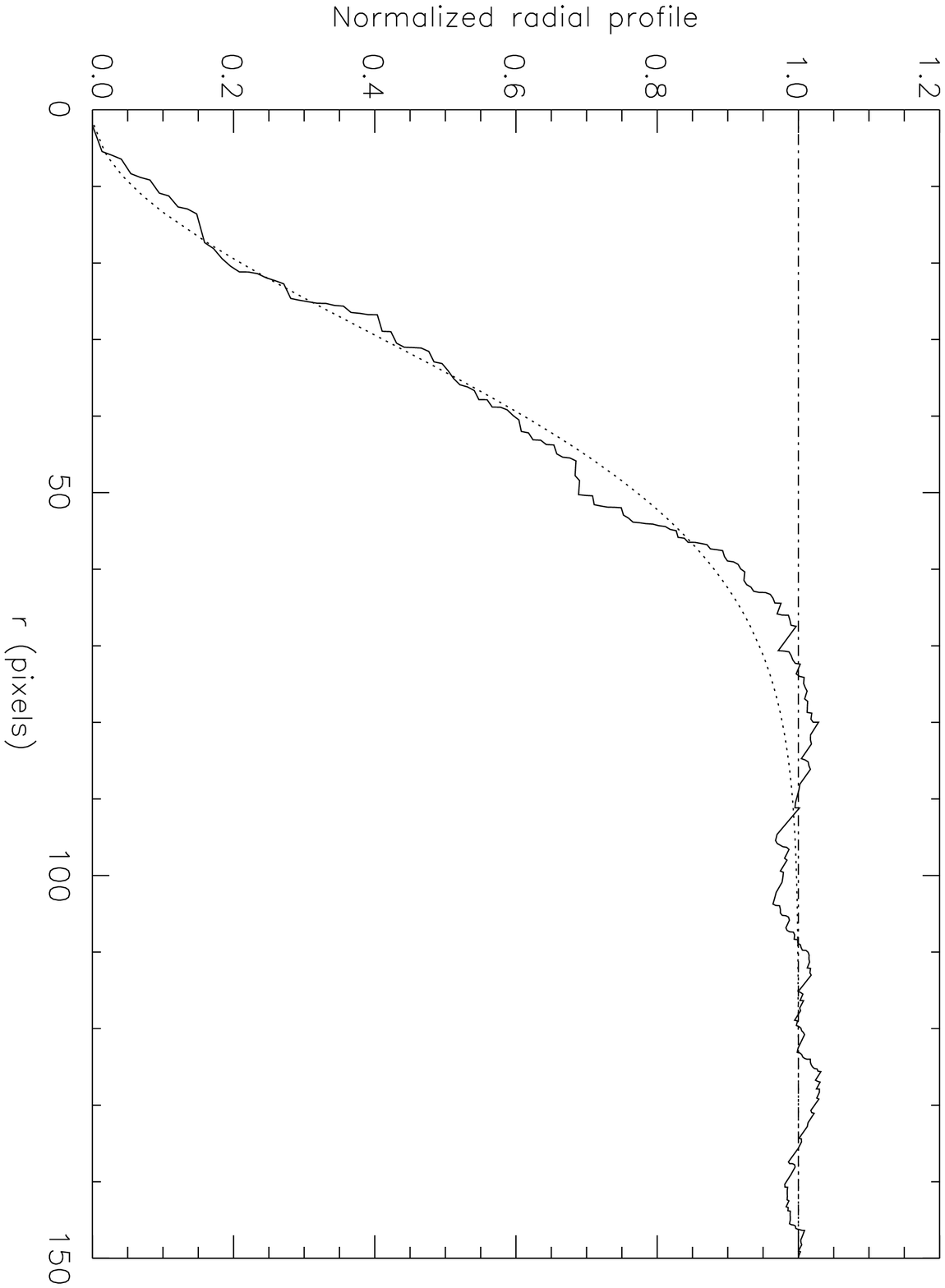}
\plotone{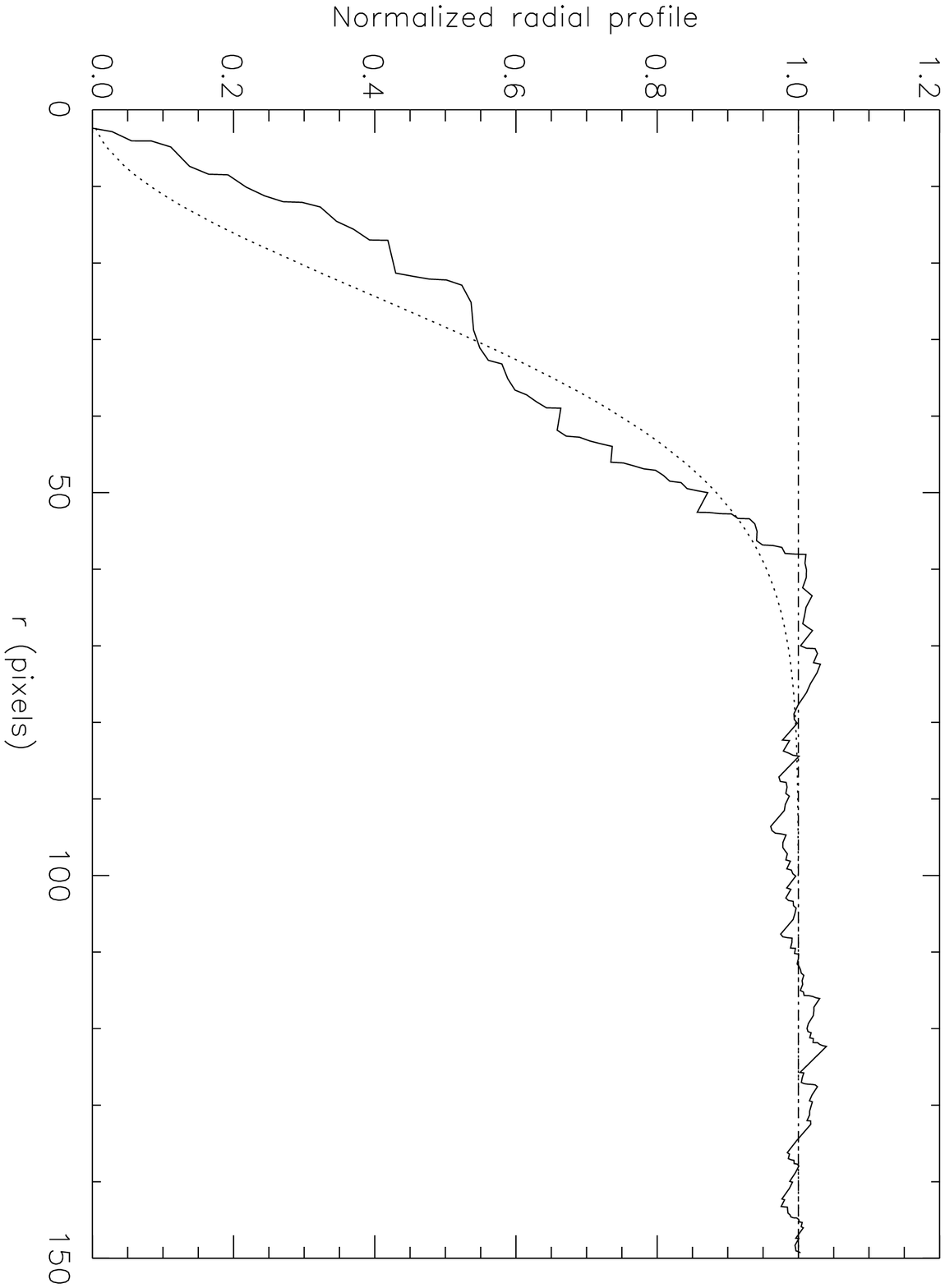}
\plotone{fig25.ps}
\plotone{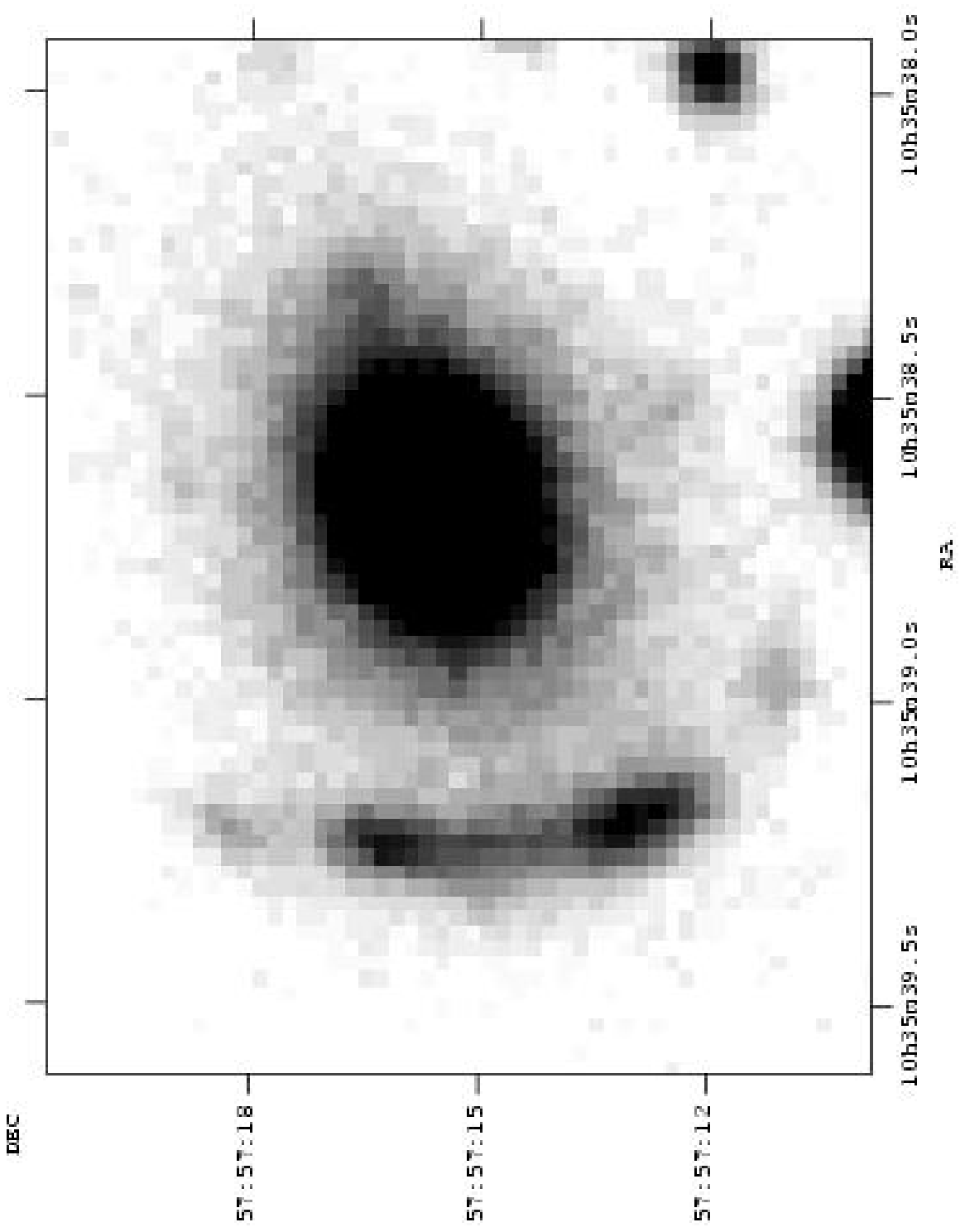}

\end{document}